\documentclass[sigconf]{acmart}

\usepackage{amsmath,amsthm,algorithm,algorithmic,bm,CJKutf8,enumitem,multirow,multicol,makecell,graphicx,pifont}

\copyrightyear{2026}
\acmYear{2026}
\setcopyright{cc}
\setcctype{by}
\acmConference[KDD '26]{Proceedings of the 32nd ACM SIGKDD Conference on Knowledge Discovery and Data Mining V.1}{August 09--13, 2026}{Jeju Island, Republic of Korea}
\acmBooktitle{Proceedings of the 32nd ACM SIGKDD Conference on Knowledge Discovery and Data Mining V.1 (KDD '26), August 09--13, 2026, Jeju Island, Republic of Korea}
\acmPrice{}
\acmDOI{10.1145/3770854.3780159}
\acmISBN{979-8-4007-2258-5/2026/08}

\begin{document}

\title{Breaking Robustness Barriers in Cognitive Diagnosis: \\ A One-Shot Neural Architecture Search Perspective}

\author{Ziwen Wang}
\affiliation{%
  \institution{School of Computer Science and Technology, Key Laboratory of Intelligent Computing and Signal Processing of Ministry of Education, Anhui University}
  \city{Hefei}
  \state{Anhui}
  \country{China}
}
\affiliation{%
  \institution{Anhui Province Key Laboratory of Intelligent Computing and Applications}
  \country{}  
}
\email{wzw12sir@gmail.com}

\author{Shangshang Yang}
\authornote{Corresponding author.}
\affiliation{%
  \institution{School of Computer Science and Technology, Anhui University}
  \city{Hefei}
  \state{Anhui}
  \country{China}
}
\affiliation{%
  \institution{Anhui Province Key Laboratory of Intelligent Computing and Applications}
  \country{}  
}
\email{yangshang0308@gmail.com}

\author{Xiaoshan Yu}
\affiliation{%
  \institution{School of Artificial Intelligence, Anhui University}
  \city{Hefei}
  \state{Anhui}
  \country{China}
}
\email{yxsleo@gmail.com}

\author{Haiping Ma}
\affiliation{%
  \institution{Institutes of Physical Science and Information Technology, Anhui University}
  \city{Hefei}
  \state{Anhui}
  \country{China}
}
\email{hpma@ahu.edu.cn}

\author{Xingyi Zhang}
\affiliation{%
  \institution{School of Computer Science and Technology, Anhui University}
  \city{Hefei}
  \state{Anhui}
  \country{China}
}
\email{xyzhanghust@gmail.com}

\renewcommand{\shortauthors}{Wang et al.}
\renewcommand{\abstractname}{ABSTRACT}
\renewcommand{\keywordsname}{KEYWORDS}
\renewcommand{\refname}{REFERENCES}

\begin{abstract}
With the advancement of network technologies, intelligent tutoring systems~(ITS) have emerged to deliver increasingly precise and tailored personalized learning services. Cognitive diagnosis~(CD) has emerged as a core research task in ITS, aiming to infer learners’ mastery of specific knowledge concepts by modeling the mapping between learning behavior data and knowledge states.
However, existing research prioritizes model performance enhancement while neglecting the pervasive noise contamination in observed response data, significantly hindering practical deployment. Furthermore, current cognitive diagnosis models~(CDMs) rely heavily on researchers' domain expertise for structural design, which fails to exhaustively explore architectural possibilities, thus leaving model architectures' full potential untapped.
To address this issue, we propose \textbf{OSCD}, an evolutionary multi-objective \underline{\textbf{O}}ne-\underline{\textbf{S}}hot neural architecture search method for \underline{\textbf{C}}ognitive \underline{\textbf{D}}iagnosis, designed to efficiently and robustly improve the model's capability in assessing learner proficiency.
Specifically, OSCD operates through two distinct stages: training and searching. During the training stage, we construct a search space encompassing diverse architectural combinations and train a weight-sharing supernet represented via the complete binary tree topology, enabling comprehensive exploration of potential architectures beyond manual design priors. In the searching stage, we formulate the optimal architecture search under heterogeneous noise scenarios as a multi-objective optimization problem~(MOP), and develop an optimization framework integrating a Pareto-optimal solution search strategy with cross-scenario performance evaluation for resolution. Extensive experiments on real-world educational datasets validate the effectiveness and robustness of the optimal architectures discovered by our OSCD model for CD tasks.
\end{abstract}




\maketitle

\section{INTRODUCTION}

\begin{figure}[t]
\centering
\includegraphics[width=\linewidth]{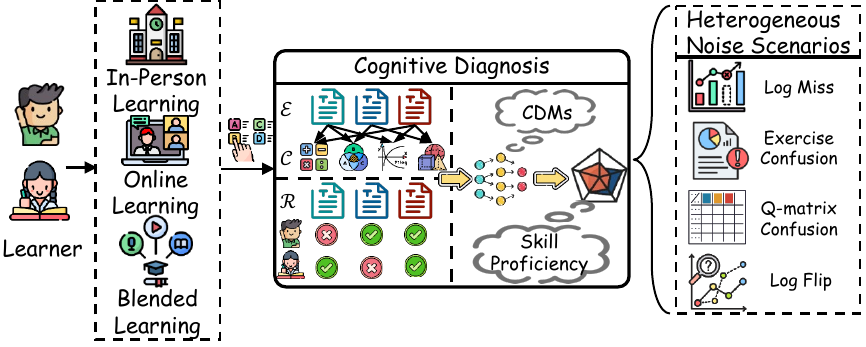}
\vspace{-3mm}
\caption[t]{The pipeline of cognitive diagnosis under heterogeneous noise scenarios.}
\label{fig:introduction1}
\vspace{-3mm}
\end{figure}

\begin{figure*}[t]
\centering
\includegraphics[width=0.9\linewidth]{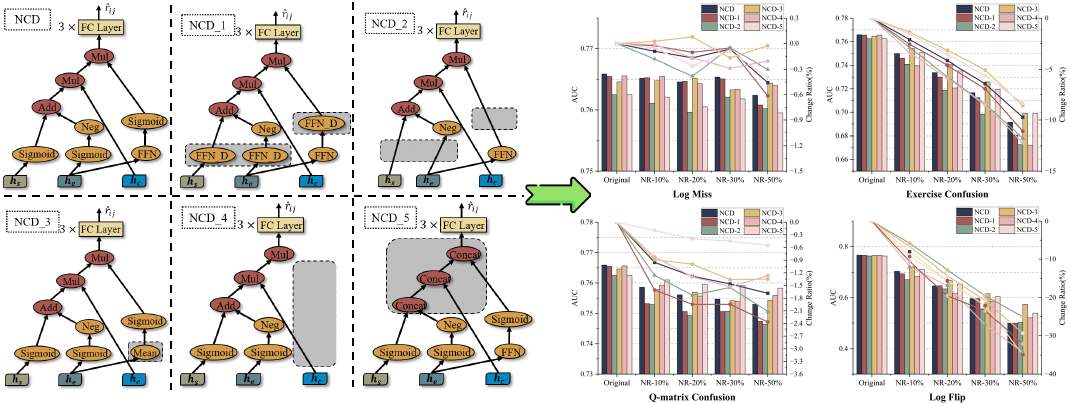}
\vspace{-3mm}
\caption[t]{Comparative performance of NCD and its Variants under different noise ratios.}
\label{fig:introduction2}
\vspace{-3mm}
\end{figure*}

The rapid advancement of network technologies has catalyzed the swift development and widespread deployment of intelligent tutoring systems~(ITS), which aim to provide personalized and adaptive learning experiences~\cite{FBKT,DRE,Disenqnet}. Within the core framework of these systems, cognitive diagnosis~(CD) serves as a fundamental component, dedicated to precisely modeling students' latent knowledge states and skill proficiencies by analyzing their interactive learning behaviors and performance data.

Cognitive diagnosis methodologies can be broadly categorized into three main paradigms: psychometric theory-based methods~\cite{DINA,IRT,MIRT}, neural network-based methods~\cite{NCD,ESRCD,PromptCD}, and graph modeling-based methods~\cite{RCD,DisenGCD,ISGCD}. The first type, representing the foundational stage of research, primarily leverages psychological measurement techniques to assess learners' latent abilities or traits. With the advent and flourishing of deep learning, researchers have increasingly turned to diverse neural network architectures to capture the complex, nonlinear student-exercise interaction patterns, thereby learning more effective representations. Furthermore, the critical importance of modeling both explicit and implicit high-order interactions among students, exercises, and knowledge concepts has gained significant recognition. This focus on relational complexity naturally motivates the introduction of graph structures into CD, leveraging their inherent strengths to model more comprehensive representations of the cognitive landscape.

Despite the remarkable progress achieved by existing CD methods, a critical limitation persists: they generally overlook the prevalent heterogeneous noise contamination in real educational environments. These approaches typically operate under the implicit assumption of relatively clean data, failing to explicitly address the adverse impacts of noisy data from various sources, including log miss, exercise confusion, $Q$-matrix confusion, and log flip, as illustrated in Figure~\ref{fig:introduction1}.
Furthermore, different model architectures exhibit varying levels of sensitivity to different noise sources, making it highly challenging to manually design a single, universally optimal structure capable of withstanding all types of heterogeneous noise contamination. 
As illustrated in Figure~\ref{fig:introduction2}, we systematically modify the widely recognized NCD model by applying various replacements and ablations, resulting in multiple architectural variants. These variants exhibit significantly divergent performance change rates across diverse heterogeneous noise environments, directly validating that the noise tolerance of CD models is highly dependent on their underlying architecture.

To this end, we proposed \textbf{OSCD}, an evolutionary multi-objective \textbf{O}ne-\textbf{S}hot neural architecture search approach for \textbf{C}ognitive \textbf{D}iagnosis to explore efficient and robust model architectures.
Specifically, OSCD operates through two sequential stages: training and searching.
In the first stage, a search space encompassing diverse architectural combinations is constructed, and a weight-sharing supernet represented via the complete binary tree topology is trained. This process enables comprehensive exploration of potential architectures that transcend manual design priors.
In the second stage, we explicitly formulate the search for the optimal architecture under heterogeneous noise scenarios as a MOP. To resolve this MOP, we develop an optimization framework that integrates a Pareto-optimal solution search strategy with cross-scenario performance evaluation.
Finally, extensive experiments conducted on real-world educational datasets validate the effectiveness and robustness of the optimal architectures discovered by OSCD for CD tasks.


\section{RELATED WORK}
\subsection{Cognitive Diagnosis}
Cognitive diagnosis, as an advanced learner modeling approach, is theoretically grounded in the intersection of modern psychometrics and cognitive science. By analyzing students' interactive learning behaviors, it can systematically deconstruct learners' knowledge states and cognitive processes.
Current research primarily focuses on manually developing sophisticated cognitive diagnostic models (CDMs) for more accurate learner modeling and assessment, which can be broadly classified into three categories: psychometric theory-based methods~\cite{DINA,IRT,MIRT}, neural network-based methods~\cite{NCD,KSCD,DZCD}, and graph modeling-based methods~\cite{RCD,DGCD,ORCDF,RDGT}. For instance, NCD\cite{NCD}, as a representative neural CDM, employs multidimensional parameters to construct fine-grained representations of students' cognitive states and exercise attributes, while leveraging neural networks to model complex interactions. In addition, RCD\cite{RCD} models student interactions and structural relationships by constructing a student-exercise-concept graph, employing graph convolutional networks (GCN\cite{GCN}) to enhance the effectiveness of representation learning. More details of CD works are provided in the \textbf{Appendix A.1.1}.

With the increasing attention to learner modeling, recent years have also witnessed critical explorations from other perspectives, including interpretability\cite{ID-CDF,KanCD}, robustness\cite{AdaRD}, and group learning\cite{HomoGCD,RIGL}. Despite the impressive results achieved by these cognitive diagnostic methods, their underlying CDM architectures often rely heavily on researchers' domain expertise and experience. Such handcrafted models are limited in their ability to comprehensively explore all possible combinations of learning behaviors, thereby constraining the generalization and further investigation of cognitive diagnosis.

\subsection{Neural Architecture Search}
As a revolutionary automated machine learning technology\cite{NASNet,NASsurvey2}, Neural Architecture Search (NAS) has evolved into a comprehensive multi-domain and multi-modal framework since its pioneering works\cite{NAS,NASsurvey1,ENASsurvey1}. Currently, NAS techniques have achieved cross-domain penetration by offering automated optimization solutions for a wide range of mainstream deep neural networks (DNNs). This technology has been successfully applied in computer vision~(CV)\cite{BigNAS}, natural language processing~(NLP)\cite{TextNAS}, speech recognition systems\cite{ASRNAS}, and graph neural networks~(GNNs)\cite{GraphNASsurvey1}. Notably, the Transformer architecture, as a universal paradigm for cross-modal applications, has also benefited from the systematic exploration enabled by NAS methods\cite{ViTAS,TransformerNASsurvey1,NAS-BERT,Darts-Conformer}. More details of NAS works are provided in the \textbf{Appendix A.1.2}.

Meanwhile, several recent pioneering studies\cite{ding2020automatic,NAS-GCD,ding2023approach} have initiated the exploration of NAS techniques in the education domain. NASCD\cite{NASCD} adopts a NAS-based approach to search for CDMs, with its primary contribution lying in the design of a revised CD search space and the definition of an interpretability-oriented search objective, enabling the effective discovery of more interpretable and efficient models. Furthermore, ENAS-KT\cite{ENAS-KT} achieves an optimal balance between input feature selection and architectural design by automatically selecting input features and formulating a novel Transformer-based search space.
However, existing research has overly emphasized the high performance and interpretability of the searched CDMs, while insufficiently considering the robustness of these models in real-world scenarios. More critically, these methods have yet to incorporate computational complexity into their design considerations, leading to severe efficiency bottlenecks when scaling to complex interaction scenarios, which significantly hampers their developmental potential.

\section{PRELIMINARY}

\subsection{Cognitive Diagnosis Task}
In the field of intelligent education, let $\mathcal{S}=\{s_1, s_2, \cdots, s_N\}$, $\mathcal{E}=\{e_1, e_2, \cdots, e_M\}$, and $\mathcal{C}=\{c_1, c_2, \cdots, c_K\}$ denote the sets of learners, exercises, and knowledge concepts, respectively, where $N$, $M$, and $K$ represent the size of each set. The relationship between exercises and knowledge concepts is represented by a $Q$-matrix, denoted as $\mathbf{Q}=\{q_{ij}\}^{M\times K}$, where $q_{ij}=1$ if exercise $e_i$ involves concept $c_j$, and $q_{ij}=0$ otherwise. The observed response data are provided in the form of the triplet set $\mathcal{R}=\{(s_i,e_j,r_{ij})|s_i\in \mathcal{S},e_j\in \mathcal{E},r_{ij}\in \{0,1\}\}$, where $r_{ij}=1$ denotes the learner $s_i$ answers the exercise $e_j$ correctly, and $r_{ij}=0$ otherwise. Naturally, the response data $\mathcal{R}$ is partitioned into training set $\mathcal{R}_\mathrm{Tr}$, validation set $\mathcal{R}_\mathrm{Val}$ and test set $\mathcal{R}_\mathrm{Te}$.
In this work, we construct multiple realistic heterogeneous noise scenarios. Specifically, we generate the perturbed validation sets $\tilde{\mathcal{R}}_\mathrm{Val}=\{\mathcal{R}_\mathrm{Val}^\Theta, \mathcal{R}_\mathrm{Val}^\Phi, \mathcal{R}_\mathrm{Val}^\Psi, \mathcal{R}_\mathrm{Val}^\Omega \}$ by applying distinct noise perturbations $\{ \Theta, \Phi, \Psi, \Omega \}$ to the original validation set $\mathcal{R}_\mathrm{Val}$ derived from response data $\mathcal{R}$. Among these, $\Theta$ corresponds to partial response log miss, where $\mathcal{R}_\mathrm{Val}^\Theta \subset \mathcal{R}_\mathrm{Val}$; $\Phi$ reflects partial confusion for response exercises, where $\mathcal{R}_\mathrm{Val}^\Phi=\{(s_i,e'_j,r_{ij})|e'_j\neq e_j,e'_j\in \mathcal{E}\}$; $\Psi$ represents the confusion of the $Q$-matrix, where $\mathbf{Q}=\{q'_{ij}\}^{M\times K}$ and $q'_{ij}\neq q_{ij}$; $\Omega$ denotes the flip of the response log, where $\mathcal{R}_\mathrm{Val}^\Omega=\{(s_i,e_j,r'_{ij})|r'_{ij}\neq r_{ij},r'_{ij}\in \{0,1\}\}$.
Typically, the noise interactions in real-world data remain implicit and unlabeled. Accordingly, the problem in this paper can be defined as follows:

\textbf{Core cognitive diagnosis task.} \textit{Given the learner-exercise response record $\mathcal{R}$ and the predefined $Q$-matrix $\mathbf{Q}$, the primary objective of cognitive diagnosis is to accurately and reliably estimate the learner’s proficiency on specific knowledge concepts.}

\textbf{Robust architecture search.} \textit{Given the training set $\mathcal{R}_\mathrm{Tr}$, the validation sets $\mathcal{R}_\mathrm{Val} \cup \tilde{\mathcal{R}}_\mathrm{Val}$ and the test set $\mathcal{R}_\mathrm{Te}$ of response records $\mathcal{R}$ along with the $Q$-matrix $\mathbf{Q}$, the primary objective of robust architecture search is to identify one or more model architectures that maintain stable and competitive performance across both original and perturbed validation sets. The searched architectures are then retrained from scratch on the training set and evaluated on the held-out test set $\mathcal{R}_\mathrm{Te}$.}

\subsection{General Methodology of CD}
For a general cognitive diagnostic task, the model accepts three main types of inputs and outputs a prediction score for student $s_i$.
In a general cognitive diagnosis framework, the model $\mathcal{F}$ takes three primary types of inputs and outputs an estimated probability $\hat{r}_{ij}$ for student $s_i$ on exercise $e_j$:
\begin{equation}
\label{eq:1}
\begin{gathered}
    \hat{r}_{ij}=\mathcal{F}(\mathbf{h}_s,\mathbf{h}_e,\mathbf{h}_c), \\
    \mathbf{h}_s = \mathbf{x}_i^s\times \mathrm{W}_s,
    \mathbf{h}_e = \mathbf{x}_j^e\times \mathrm{W}_e,
    \mathbf{h}_c = \mathbf{x}_j^e\times \mathrm{Q},\\
\end{gathered}
\end{equation}
where $\mathbf{h}_s \in \mathbb{R}^{1\times d}$ is the learner-related feature embedding, $\mathbf{h}_e \in \mathbb{R}^{1\times d}$ is the exercise-related feature embedding, $\mathbf{h}_c \in \mathbb{R}^{1\times K}$ is the concept-related feature embedding, $d$ denotes the embedding dimension, $\mathbf{x}_i^s \in \{0,1\}^{1\times N}$ denotes the one-hot vector of student $s_i$, $\mathbf{x}_i^e \in \{0,1\}^{1\times M}$ denotes the one-hot vector of exercise $e_j$, $\mathrm{W}_s$ and $\mathrm{W}_e$ are trainable matrices.

\begin{table}[t]
\renewcommand{\arraystretch}{0.1}       
\centering
\caption{Detailed explanation of operators.}
\vspace{-3mm}
\label{tab:operators}
\resizebox{\linewidth}{!}{
\begin{tabular}{c|cccc}
\toprule
Notation & Syntax & meaning & Output Shape & Lipschitz Condition                   \\ 
\midrule
Zero & $\backslash$ & \makecell[c]{Placeholder \\ (Input or Operation)} & $\backslash$ & $\backslash$   \\
\midrule
Iden & $x$ & Identity & same & \ding{52} \\
Neg & $-x$ & Negative & same & \ding{52}   \\
Abs & $|x|$ & Absolute value & same & \ding{52}     \\
Inv & $1/(x+1e^{-6})$ & Inverse & same & \ding{56} \\ 
Square & $x^2$ & Square & same & \ding{56} \\ 
Sqrt & $sign(x)\cdot \sqrt{|x|+1e^{-6}}$ & Square root & same & \ding{56} \\ 
Tanh & $tanh(x)$ & Tanh function & same & \ding{52} \\ 
Sigmoid & $sigmoid(x)$ & Sigmoid function & same & \ding{52} \\ 
Softplus & $softplus(x)$ & Softplus function & same & \ding{52} \\ 
Sum & $\sum_i^d x_i$ & Sum of the vector & $\mathbb{R}^1$ & \ding{52} \\ 
Mean & $(\sum_i^d x_i)/d$ & Mean of the vector & $\mathbb{R}^1$ & \ding{52} \\ 
FFN & \makecell[c]{$x_i\times \mathrm{W}_{ffn}$, \\ $\mathrm{W}_{ffn}\in \mathbb{R}^{d\times 1}$} & \makecell[c]{An FC layer mapping \\ a vector to a scalar} & $\mathbb{R}^1$ & \ding{52} \\ 
FFN\_D & \makecell[c]{$x_i\times \mathrm{W}_{ffnd}$, \\ $\mathrm{W}_{ffnd}\in \mathbb{R}^{d\times d}$} & \makecell[c]{An FC layer mapping \\ a vector to a vector} & $\mathbb{R}^{1\times d}$ & \ding{52} \\
\midrule
Add & $x+y$ & Addition & $max(x,y)$ & \ding{52} \\ 
Mul & $x\cdot y$ or $x\odot y$ & Multiplication & $max(x,y)$ & \ding{56} \\ 
Concat & \makecell[c]{$[x||y]\times \mathrm{W}_{concat}$, \\ $\mathrm{W}_{concat}\in \mathbb{R}^{2d\times d}$} & \makecell[c]{Concatenate and map \\ two vectors} & $\mathbb{R}^{1\times d}$ & \ding{52} \\ 
\bottomrule
\end{tabular}
}
\vspace{-3mm}
\end{table}

\section{METHODOLOGY}
\subsection{Overview of The Proposed OSCD}
As illustrated in Figure~\ref{fig:ea} (see Algorithm 1 for stepwise summary in \textbf{Appendix A.2}), our proposed OSCD model comprises two core stages: training and searching. During the training stage, we first design a search space for CD tasks consisting of two components: the first integrates three types of input nodes and $17$ internal computational nodes, while the second dynamically selects the final prediction layer based on the output dimension from the preceding part. We then construct and train a supernet based on this search space. To fully explore the combinatorial potential of models, we refine the search space into a complete binary tree topology representation. This structured representation significantly facilitates the development and evaluation of all possible model combinations.

Upon completion of supernet training, the search stage commences by initializing a population of $Pop$ individuals. Some individuals are initialized using existing high-performance manually-designed architectures to leverage their empirical value and accelerate search convergence, while the remaining individuals are randomly generated from the search space. Subsequently, parent individuals are selected through standard binary tournament selection to form a mating pool, followed by the application of single-point crossover and bit-wise mutation to generate offspring populations. Using the trained supernet weights, we efficiently evaluate the performance of all newly generated offspring architectures. Then, employing the environmental selection mechanism of NSGA-II~\cite{NSGA2}, we filter individuals with superior fitness from the combined parent and offspring populations to form a new generation. This search cycle 
iterates until reaching the maximal generation count, ultimately outputting all non-dominated individuals in the final population as the Pareto-optimal solution set.

\subsection{Training Stage}
\subsubsection{Search space}
\begin{figure}[t]
\centering
\includegraphics[width=0.8\linewidth]{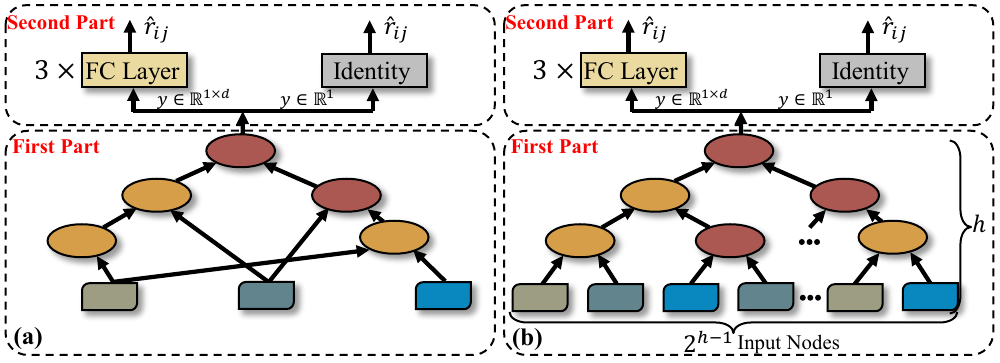}
\vspace{-3mm}
\caption[t]{The generalized search space(a) and the fine-tuned search space(b).}
\label{fig:search_space1}
\vspace{-3mm}
\end{figure}

As mentioned earlier, the search spaces employed by existing NAS methods are task-specific and cannot be directly applied to the CD task\cite{NASsurvey1,NASsurvey2}.
Therefore, we systematically investigate and summarize the architectural characteristics of existing CDMs. They can be uniformly abstracted into a general paradigm comprising multiple input nodes, multiple internal nodes, and one output node, as illustrated in Figure~\ref{fig:search_space1}(a).

To enable the general model to represent as many CDMs as possible, we systematically integrated $17$ categories of representative feature transformation operations, which are structurally embedded in the model's internal computational nodes to support general representations of diverse CDMs. Table~\ref{tab:operators} presents the details of these $17$ operators and their compliance with the Lipschitz condition, which are primarily categorized into three types: \textbf{a placeholder operation}(applicable to both input and operation), \textbf{unary operations}~(one input), and \textbf{binary operations}~(two inputs).
It should be specially noted that we have categorized five operators: $Sum$, $Mean$, $FFN$, $FFN\_D$, and $Concat$, collectively termed as \textbf{dimension-sensitive operators}. The key characteristic of these operators is that they require $d$-dimension vectors as input to execute their specific computational logic.
Specifically, to address the critical challenge of potential partial path missing in tree-like topological structures within the search space, we innovatively introduce a placeholder operation~($Zero$) mechanism. This mechanism maintains the continuity of feature transmission in incomplete binary tree structures by constructing virtual paths, ultimately enabling complete cognitive state derivation from leaf nodes to the root node. Furthermore, we conduct rigorous verification of Lipschitz continuity for these $17$ operator categories, precisely determining the upper bounds of their Lipschitz constants. This mathematical property directly characterizes the output stability of nodes under heterogeneous noise perturbation. Additional details regarding the operators and proofs of the Lipschitz condition are provided in the \textbf{Appendix A.3} due to space constraints.

As shown in Table~\ref{tab:operators}, all operations output either a scalar or a $d$-dimension vector, indicating that the general model must ultimately produce either a scalar $y$ or a $d$-dimension vector $\mathbf{y}$. To accomplish the prediction task, we need to specify corresponding computational processes for both scalar and $d$-dimension vector outputs, as detailed below:
\begin{equation}
    \hat{r}_{ij}=\left\{
        \begin{aligned}
         & y,~if~y\in \mathbb{R}^1 \\
         \mathrm{FC}_3( & \mathrm{FC}_2(\mathrm{FC}_1(\mathbf{y})),~if~y\in \mathbb{R}^{1\times d}
        \end{aligned}
        \right.,
\end{equation}
where $\mathrm{FC}_1(\cdot)$, $\mathrm{FC}_2(\cdot)$, and $\mathrm{FC}_3(\cdot)$ are three Fully Connected(FC) layers with output dimensions $512$, $256$, $1$, respectively. It can also maintain certain monotonicity properties, which has gained recognition from numerous researchers and has been widely adopted in most CD studies\cite{NCD,RCD}.

Therefore, the variable components in the general model constitute the search space for CD, primarily consisting of two parts. The first part mainly includes input nodes and internal computational nodes that form the primary computational path. The second part employs an identity operation and a three-layer FC network to accomplish the prediction task. As illustrated in Figure~\ref{fig:search_space1}(a), the first part contains up to three types of input nodes (represented by rounded rectangles) and $17$ internal computational nodes (represented by ovals, with red indicating binary operations and yellow indicating unary operations). The second part primarily consists of an identity operation (represented by a gray rectangle) and a three-layer FC network (represented by yellow rectangles) to obtain the final predicted value $\hat{r}_{ij}$.

\begin{figure*}
\centering
\includegraphics[width=0.95\linewidth]{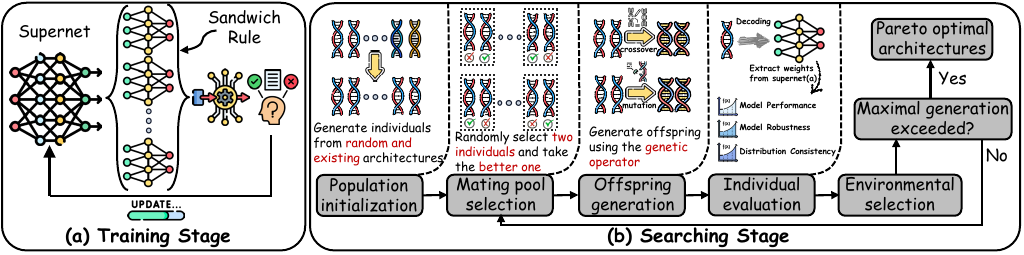}
\vspace{-3mm}
\caption[t]{The overview of our proposed OSCD.}
\label{fig:ea}
\vspace{-3mm}
\end{figure*}

\begin{figure}[t]
\centering
\includegraphics[width=0.8\linewidth]{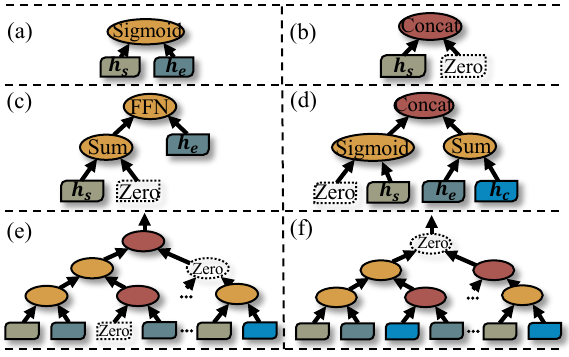}
\vspace{-3mm}
\caption[t]{Key challenges (a-f) in supernet training.}
\label{fig:search_space2}
\vspace{-3mm}
\end{figure}

\subsubsection{Supernet}
However, while the search space incorporating diverse input nodes and internal computational nodes offers expressive flexibility, it suffers from a critical flaw: the combinatorial explosion creates an excessively large search space that imposes prohibitively high computational complexity on traditional NAS methods. Specifically, the independent training and evaluation processes for numerous substructures are computationally intensive, and the discrete search mechanism struggles to efficiently converge to high-performance regions. 

To overcome computational efficiency bottlenecks and accelerate optimization convergence, we innovatively introduce a supernet architecture. Constructed upon the aforementioned search space, this architecture employs weight-sharing mechanisms and layered sandwich training rules to achieve synchronous representation and evaluation of massive subnets, substantially reducing the time complexity of traditional NAS methods and establishing an efficient computational foundation for the subsequent MOP. Based on the above, we employ a tree-based topological structure to encode CDMs, as shown in Figure~\ref{fig:search_space1}(b). Specifically, we construct a full binary tree topology with height $h$, where all input nodes serve as leaf nodes, intermediate nodes function as non-leaf nodes, and the output node corresponds to the root node. This results in a full binary tree topology consisting of $2^{h-1}$ leaf nodes and $2^{h-1}-1$ non-leaf nodes, representing $2^{h-1}$ input nodes ($Zero$, $\mathbf{h}_s$,$\mathbf{h}_e$, and $\mathbf{h}_c$) and $2^{h-1}-1$ operation nodes ($17$ operators in Table~\ref{tab:operators}), respectively. Naturally, for each architecture, its $i$-th operation node can be represented as $node_i=\{input_i^L, input_i^R, operator_i\}$, where $input_i^L$, $input_i^R$, and $operator_i$ respectively denote the computational result of the left subtree, the right subtree, and the currently selected operator. The prediction result $\hat{r}_{ij}$ is ultimately generated through recursive computation of the feature propagation process from leaf nodes to the root node. 

However, reality proves less ideal as this full binary tree representation paradigm encounters significant challenges when generalizing to complex CDMs:
(1) \textbf{Dimension Incompatibility:} This primarily manifests as matching errors between input nodes and operators, which can be categorized into four typical scenarios:
(a) \textit{Unary operator mismatch}: as illustrated in Figure~\ref{fig:search_space2}(a), when a unary operation ($Sigmoid$) is incorrectly connected to two inputs $( \mathbf{h}_s, \mathbf{h}_e)$, we enforce computation using the left input: $sigmoid(\mathbf{h_s})$.
(b) \textit{Binary operator deficiency}: as shown in Figure~\ref{fig:search_space2}(b), when a binary operation ($Concat$) receives only a valid left input $\mathbf{h}_s$ while the right input is an invalid placeholder $Zero$, we pad the right dimension with an all-ones tensor: $Concat(\mathbf{h}_s,\mathbf{1}^{1\times d})$.
(c) \textit{Dimension-sensitive operator left input error}: as depicted in Figure~\ref{fig:search_space2}(c), when a dimension-sensitive operator ($FFN$) receives a scalar $Sum(\mathbf{h}_s)$ as left input but requires a $d$-dimensional vector, we directly substitute it with a dimensionless operator to avoid dimensional conflicts: $Sigmoid(Sum(\mathbf{h}_s))$.
(d) \textit{Dimension-sensitive operator right input error}: as can be seen from Figure~\ref{fig:search_space2}(d), when the left input $Sigmoid(\mathbf{h}_s)\in \mathbb{R}^{1\times d}$ but the right input is a scalar $Sum(\mathbf{h}_e)\in \mathbb{R}^1$ (in $Concat$ operations), we expand the right input dimension using an all-ones tensor: $Concat(Sigmoid(\mathbf{h}_s), [Sum(\mathbf{h}_e)|| \mathbf{1}^{1\times (d-1)}])$.
Detailed explanations are provided in \textbf{Appendix A.4}.

(2) \textbf{Path Discontinuity:} A significant portion of computational paths originating from leaf nodes fail to reach the root node, as shown in Figure~\ref{fig:search_space2}(e). This occurs because the computational graph contains numerous $Zero$ placeholders acting as operators. To address this, we preserve the active computational flows along the paths. $Zero$ placeholders are tagged accordingly, and these tags are subsequently used to validate path viability and determine the final output.

(3) \textbf{Structural Collapse:} As illustrated in Figure~\ref{fig:search_space2}(f), when computational paths cannot extend to the root node, or when both the root node and all its child nodes are $Zero$ placeholders, the model loses its effective predictive capability. To address structural collapse that disables prediction functionality, we introduce a dynamic fallback mechanism. When it detects either that no computation path originating from leaf nodes can reach the root, or the root node itself is a $Zero$ placeholder, it automatically activates a contingency procedure. This procedure traces back to the highest-level valid computational node proximal to the root and adopts its output as the final prediction. This mechanism elevates system fault tolerance by ensuring graceful degradation under structural collapse, thereby preserving continuous prediction capability without compromising the core computational paradigm.

Building upon the solutions to the aforementioned issues and inspired by the sandwich rule~\cite{BigNAS}, we propose employing three types of subnets in each training batch of the supernet: the minimal subnet (with the smallest parameters representing the lower bound), the maximal subnet (with the largest parameters representing the upper bound), and multiple randomly sampled subnets (typically $3$ samples). Consequently, this approach not only emphasizes training boundaries but also introduces randomness during supernet training, achieving a balance between extreme search and fair exploration of the search space, thereby enabling the trained supernet weights to better facilitate subsequent subnet performance evaluation. The \textbf{Appendix A.5} presents the specific details.

\begin{table*}[t]
\centering
\caption{The prediction performance comparison between the architectures discovered by our proposed OSCD and baselines on the ASSIST09 and SLP dataset. $\uparrow$~($\downarrow$) means the higher~(lower) score the better performance. \textbf{Bold}: the best, \underline{Underline}: the runner-up.}
\vspace{-6mm}
\label{tab:main_experiment}
\caption*{(a) Results on the ASSIST09 dataset.}
\vspace{-5mm}
\resizebox{1.0\linewidth}{!}{
    \begin{tabular}{c|c|ccc|ccc|ccc|ccc}
    \toprule
    \multicolumn{2}{c|}{Train/Val/Test} & \multicolumn{3}{c|}{50\%/20\%/30\%} & \multicolumn{3}{c|}{60\%/20\%/20\%} & \multicolumn{3}{c|}{70\%/10\%/20\%} & \multicolumn{3}{c}{80\%/10\%/10\%} \\
    \midrule
    \multicolumn{2}{c|}{Metrics} & ACC($\uparrow$)   & AUC($\uparrow$)   & RMSE($\downarrow$)  & ACC($\uparrow$)   & AUC($\uparrow$)   & RMSE($\downarrow$)  & ACC($\uparrow$)   & AUC($\uparrow$)   & RMSE($\downarrow$)  & ACC($\uparrow$)   & AUC($\uparrow$)   & RMSE($\downarrow$) \\
    \midrule
    \multicolumn{2}{c|}{DINA} & 0.6383  & 0.6902  & 0.4927  & 0.6528  & 0.7036  & 0.4856  & 0.6628  & 0.7150  & 0.4798  & 0.6759  & 0.7282  & 0.4738  \\
    \multicolumn{2}{c|}{IRT} & 0.6901  & 0.6889  & 0.4742  & 0.6960  & 0.7140  & 0.4505  & 0.7103  & 0.7187  & 0.4535  & 0.7160  & 0.7271  & 0.4488  \\
    \multicolumn{2}{c|}{MIRT} & 0.6983  & 0.7170  & 0.4824  & 0.7105  & 0.7219  & 0.4694  & 0.7147  & 0.7405  & 0.4586  & 0.7175  & 0.7480  & 0.4517  \\
    \multicolumn{2}{c|}{MF} & 0.7034  & 0.7137  & 0.4768  & 0.7089  & 0.7239  & 0.4624  & 0.7155  & 0.7361  & 0.4601  & 0.7166  & 0.7403  & 0.4578  \\
    \multicolumn{2}{c|}{NCD} & 0.7130  & 0.7298  & 0.4712  & 0.7216  & 0.7421  & 0.4474  & 0.7225  & 0.7499  & 0.4417  & 0.7270  & 0.7579  & 0.4379  \\
    \multicolumn{2}{c|}{RCD} & 0.7142  & 0.7333  & 0.4687  & 0.7259  & 0.7538  & 0.4396  & 0.7319  & 0.7585  & 0.4356  & 0.7331  & 0.7630  & 0.4343  \\
    \multicolumn{2}{c|}{KSCD} & 0.7190  & 0.7425  & 0.4451  & 0.7278  & 0.7537  & 0.4363  & 0.7288  & 0.7594  & 0.4323  & 0.7331  & 0.7662  & 0.4295  \\
    \multicolumn{2}{c|}{KaNCD} & 0.7143  & 0.7314  & 0.4586  & 0.7175  & 0.7442  & 0.4536  & 0.7250  & 0.7552  & 0.4440  & 0.7325  & 0.7623  & 0.4350  \\
    \multicolumn{2}{c|}{ReliCD} & 0.7202  & 0.7414  & 0.4377  & 0.7270  & 0.7513  & 0.4315  & 0.7317  & 0.7585  & 0.4304  & \textbf{0.7378 } & 0.7670  & 0.4259  \\
    \multicolumn{2}{c|}{ORCDF} & 0.7193  & 0.7397  & 0.4546  & 0.7243  & 0.7493  & 0.4461  & 0.7298  & 0.7554  & 0.4366  & 0.7281  & 0.7611  & 0.4359  \\
    \midrule
    \multicolumn{1}{c|}{\multirow{3}[2]{*}{EMO-NAS-CD}} & A1    & 0.7191  & 0.7469  & 0.4369  & 0.7251  & 0.7554  & 0.4306  & 0.7277  & 0.7601  & 0.4285  & 0.7293  & 0.7663  & 0.4285  \\
          & A2    & \textbf{0.7257 } & \textbf{0.7543 } & \textbf{0.4295 } & 0.7288  & \textbf{0.7610 } & \underline{0.4273}  & 0.7304  & 0.7658  & \textbf{0.4245 } & 0.7250  & \textbf{0.7745 } & 0.4797  \\
          & A3    & 0.7172  & 0.7434  & 0.4378  & 0.7283  & 0.7579  & 0.4365  & 0.7316  & 0.7626  & 0.4260  & 0.7239  & 0.7665  & 0.4240  \\
    \midrule
    \multicolumn{1}{c|}{\multirow{5}[2]{*}{OSCD}} & AA1   & 0.7228  & \underline{0.7523}  & 0.4363  & \textbf{0.7303 } & \underline{0.7605}  & \textbf{0.4267 } & \textbf{0.7341 } & \textbf{0.7661 } & 0.4254  & 0.7341  & 0.7710  & 0.4234  \\
          & AA2   & 0.7231  & 0.7518  & 0.4362  & 0.7284  & 0.7598  & 0.4296  & 0.7314  & 0.7653  & 0.4275  & 0.7334  & \underline{0.7721}  & 0.4252  \\
          & AA3   & \underline{0.7242}  & 0.7505  & 0.4349  & 0.7281  & 0.7594  & 0.4304  & \underline{0.7336}  & 0.7643  & 0.4259  & 0.7335  & 0.7720  & \textbf{0.4223 } \\
          & AA4   & 0.7211  & 0.7511  & \underline{0.4341}  & 0.7287  & 0.7588  & 0.4305  & 0.7327  & \underline{0.7655}  & 0.4253  & \underline{0.7345}  & 0.7709  & \underline{0.4233}  \\
          & AA5   & 0.7177  & 0.7507  & 0.4345  & \underline{0.7299}  & 0.7592  & 0.4306  & 0.7311  & \textbf{0.7661 } & \underline{0.4250}  & 0.7335  & 0.7713  & 0.4248  \\
    \bottomrule
    \end{tabular}%
}

\caption*{(b) Results on the SLP-Math dataset.}
\vspace{-5mm}
\resizebox{1.0\linewidth}{!}{
    \begin{tabular}{c|c|ccc|ccc|ccc|ccc}
    \toprule
    \multicolumn{2}{c|}{Train/Val/Test} & \multicolumn{3}{c|}{50\%/20\%/30\%} & \multicolumn{3}{c|}{60\%/20\%/20\%} & \multicolumn{3}{c|}{70\%/10\%/20\%} & \multicolumn{3}{c}{80\%/10\%/10\%} \\
    \midrule
    \multicolumn{2}{c|}{Metrics} & ACC($\uparrow$)   & AUC($\uparrow$)   & RMSE($\downarrow$)  & ACC($\uparrow$)   & AUC($\uparrow$)   & RMSE($\downarrow$)  & ACC($\uparrow$)   & AUC($\uparrow$)   & RMSE($\downarrow$)  & ACC($\uparrow$)   & AUC($\uparrow$)   & RMSE($\downarrow$) \\
    \midrule
    \multicolumn{2}{c|}{DINA} & 0.6378  & 0.6723  & 0.5001  & 0.6156  & 0.6573  & 0.5126  & 0.6480  & 0.7112  & 0.4865  & 0.6102  & 0.6722  & 0.5109  \\
    \multicolumn{2}{c|}{IRT} & 0.6953  & 0.7278  & 0.4843  & 0.7283  & 0.7586  & 0.4593  & 0.7487  & 0.7955  & 0.4395  & 0.6968  & 0.7415  & 0.4662  \\
    \multicolumn{2}{c|}{MIRT} & 0.7660  & 0.8231  & 0.4115  & 0.7764  & 0.8364  & 0.4057  & 0.7816  & 0.8435  & 0.3977  & 0.7762  & 0.8451  & 0.3994  \\
    \multicolumn{2}{c|}{MF} & 0.7720  & 0.8267  & 0.4134  & 0.7699  & 0.8406  & 0.4021  & 0.7863  & 0.8525  & 0.3942  & 0.7762  & 0.8500  & 0.3944  \\
    \multicolumn{2}{c|}{NCD} & 0.7683  & 0.8333  & 0.4073  & 0.7791  & 0.8452  & 0.4007  & 0.7830  & 0.8514  & 0.3950  & 0.7818  & 0.8519  & 0.3974  \\
    \multicolumn{2}{c|}{RCD} & 0.7696  & 0.8303  & 0.4081  & 0.7749  & 0.8419  & 0.3955  & 0.7802  & 0.8511  & 0.3946  & 0.7795  & 0.8504  & 0.3976  \\
    \multicolumn{2}{c|}{KSCD} & \underline{0.7829}  & 0.8426  & \underline{0.3930}  & \underline{0.7863}  & 0.8483  & \textbf{0.3901 } & \textbf{0.7943 } & 0.8575  & \underline{0.3857}  & 0.7722  & 0.8503  & 0.3965  \\
    \multicolumn{2}{c|}{KaNCD} & 0.7694  & 0.8302  & 0.4085  & 0.7744  & 0.8415  & 0.4062  & 0.7797  & 0.8503  & 0.3939  & 0.7786  & 0.8509  & 0.3970  \\
    \multicolumn{2}{c|}{ReliCD} & 0.7710  & 0.8340  & 0.4010  & 0.7843  & 0.8475  & \underline{0.3925}  & 0.7830  & 0.8523  & 0.3907  & 0.7794  & 0.8520  & 0.3926  \\
    \multicolumn{2}{c|}{ORCDF} & 0.7687  & 0.8338  & 0.4068  & 0.7799  & 0.8461  & 0.4001  & 0.7845  & 0.8530  & 0.3937  & 0.7827  & 0.8521  & 0.3969  \\
    \midrule
    \multicolumn{1}{c|}{\multirow{3}[2]{*}{EMO-NAS-CD}} & S1    & 0.7702  & 0.8276  & 0.4145  & 0.7740  & 0.8386  & 0.4041  & 0.7905  & 0.8535  & 0.3919  & 0.7786  & 0.8505  & 0.3953  \\
          & S2    & 0.7699  & 0.8295  & 0.4134  & 0.7679  & 0.8409  & 0.4127  & 0.7755  & 0.8488  & 0.4006  & 0.7754  & 0.8452  & 0.4047  \\
          & S3    & 0.7293  & 0.7800  & 0.4534  & 0.7447  & 0.7973  & 0.4403  & 0.7411  & 0.8002  & 0.4443  & 0.7329  & 0.7894  & 0.4537  \\
    \midrule
    \multicolumn{1}{c|}{\multirow{5}[2]{*}{OSCD}} & SA1   & 0.7807  & \underline{0.8442}  & 0.3973  & 0.7849  & 0.8494  & 0.3930  & 0.7811  & \textbf{0.8626 } & 0.3969  & 0.7866  & \underline{0.8537}  & 0.3930  \\
          & SA2   & 0.7797  & 0.8433  & 0.3973  & \textbf{0.7873 } & 0.8503  & 0.3935  & 0.7887  & \underline{0.8622}  & 0.3860  & \textbf{0.7906 } & 0.8527  & \underline{0.3923}  \\
          & SA3   & 0.7823  & 0.8428  & 0.3975  & 0.7825  & 0.8483  & 0.3954  & 0.7868  & 0.8618  & 0.3873  & 0.7818  & 0.8531  & 0.3936  \\
          & SA4   & 0.7781  & \underline{0.8442}  & 0.3975  & 0.7853  & \underline{0.8506}  & 0.3938  & 0.7891  & 0.8603  & 0.3866  & \underline{0.7874}  & \textbf{0.8538 } & 0.3933  \\
          & SA5   & \textbf{0.7868 } & \textbf{0.8455 } & \textbf{0.3910 } & 0.7767  & \textbf{0.8516 } & 0.3927  & 0.7929  & 0.8610  & \textbf{0.3842 } & \underline{0.7874}  & 0.8526  & \textbf{0.3907 } \\
    \bottomrule
    \end{tabular}%

}
\end{table*}

\begin{table*}[t]
\centering
\caption{The experimental results~(AUC change ratio \%) for learner performance prediction on the SLP-Math dataset under four heterogeneous noise scenarios with increasing noise ratios (10\%, 20\%, 30\%, 50\%). Avg. denotes the average of the absolute values of all AUC change ratios. The symbols '+' and '–' indicate that the average AUC performance of the five architectures discovered by the proposed OSCD is significantly better or significantly worse than that of other baselines, respectively (with a significance level of 0.05).}
\vspace{-3mm}
\label{tab:robustness_study}
\resizebox{\linewidth}{!}{
    \begin{tabular}{cc|ccccccccccccccccc}
    \toprule
    \multicolumn{2}{c|}{Noise Scenario} & \multicolumn{4}{c|}{Log Miss} & \multicolumn{4}{c|}{Exercise Confusion} & \multicolumn{4}{c|}{Q-matrix Confusion} & \multicolumn{4}{c|}{Log Flip} & \multirow{2}[4]{*}{Avg.} \\
\cmidrule{1-18}    \multicolumn{2}{c|}{Noise Ratio} & 10\%  & 20\%  & 30\%  & \multicolumn{1}{c|}{50\%} & 10\%  & 20\%  & 30\%  & \multicolumn{1}{c|}{50\%} & 10\%  & 20\%  & 30\%  & \multicolumn{1}{c|}{50\%} & 10\%  & 20\%  & 30\%  & \multicolumn{1}{c|}{50\%} &  \\
    \midrule
    \multicolumn{2}{c|}{DINA} & -0.394  & -2.826  & -5.146  & \multicolumn{1}{c|}{2.095 } & -2.953  & -9.716  & -14.679  & \multicolumn{1}{c|}{-8.127 } & -1.167  & -2.826  & -4.612  & \multicolumn{1}{c|}{4.345 } & -2.390  & -11.600  & -21.780  & \multicolumn{1}{c|}{-31.594 } & 7.891  \\
    \multicolumn{2}{c|}{IRT} & -0.528  & -4.965  & -2.087  & \multicolumn{1}{c|}{3.759 } & -2.967  & -10.409  & -9.918  & \multicolumn{1}{c|}{-6.084 } & 0.000  & -5.481  & -1.332  & \multicolumn{1}{c|}{2.841 } & -7.731  & -17.876  & -24.073  & \multicolumn{1}{c|}{-35.236 } & 8.455  \\
    \multicolumn{2}{c|}{MIRT} & -0.285  & 0.688  & -0.771  & \multicolumn{1}{c|}{0.581 } & -1.695  & -3.142  & -4.126  & \multicolumn{1}{c|}{-9.152 } & -2.217  & -2.312  & -2.833  & \multicolumn{1}{c|}{-2.644 } & -9.176  & -16.811  & -25.809  & \multicolumn{1}{c|}{-38.400 } & 7.540  \\
    \multicolumn{2}{c|}{MF} & -0.481  & 0.411  & -0.352  & \multicolumn{1}{c|}{1.455 } & -1.443  & -1.713  & -3.531  & \multicolumn{1}{c|}{-5.384 } & 0.000  & -0.094  & 0.469  & \multicolumn{1}{c|}{0.246 } & -9.666  & -24.587  & -26.463  & \multicolumn{1}{c|}{-37.490 } & 7.111  \\
    \multicolumn{2}{c|}{NCD} & -0.329  & 0.975  & -0.904  & \multicolumn{1}{c|}{1.374 } & -1.268  & -1.621  & -3.406  & \multicolumn{1}{c|}{-6.225 } & -0.329  & 0.376  & -0.247  & \multicolumn{1}{c|}{-0.035 } & -9.326  & -24.090  & -26.768  & \multicolumn{1}{c|}{-37.667 } & 7.184  \\
    \multicolumn{2}{c|}{RCD} & -0.423  & 0.235  & -1.292  & \multicolumn{1}{c|}{1.974 } & -1.057  & -1.845  & -2.996  & \multicolumn{1}{c|}{-5.851 } & -0.153  & 1.492  & 1.739  & \multicolumn{1}{c|}{0.928 } & -9.482  & -24.968  & -27.212  & \multicolumn{1}{c|}{-37.892 } & 7.471  \\
    \multicolumn{2}{c|}{KSCD} & -0.327  & 0.513  & -1.096  & \multicolumn{1}{c|}{1.061 } & -1.329  & -1.574  & -2.869  & \multicolumn{1}{c|}{-5.376 } & 0.140  & 0.070  & 0.187  & \multicolumn{1}{c|}{0.303 } & -9.458  & -17.679  & -27.207  & \multicolumn{1}{c|}{-37.878 } & 6.692  \\
    \multicolumn{2}{c|}{KaNCD} & -0.823  & 0.835  & -0.576  & \multicolumn{1}{c|}{0.165 } & -2.529  & -3.316  & -5.704  & \multicolumn{1}{c|}{-12.972 } & -1.411  & -0.776  & 0.129  & \multicolumn{1}{c|}{-0.859 } & -9.032  & -16.935  & -26.355  & \multicolumn{1}{c|}{-39.857 } & 7.642  \\
    \multicolumn{2}{c|}{ReliCD} & -0.563  & 0.669  & -1.150  & \multicolumn{1}{c|}{0.927 } & -1.678  & -2.088  & -3.907  & \multicolumn{1}{c|}{-6.817 } & -0.739  & 0.023  & -0.505  & \multicolumn{1}{c|}{-0.282 } & -9.644  & -23.830  & -26.669  & \multicolumn{1}{c|}{-37.569 } & 7.316  \\
    \multicolumn{2}{c|}{ORCDF} & -0.563  & 0.410  & -1.032  & \multicolumn{1}{c|}{-0.340 } & -3.130  & -3.869  & -6.202  & \multicolumn{1}{c|}{-13.423 } & -1.876  & -1.301  & -0.352  & \multicolumn{1}{c|}{-1.348 } & -9.496  & -17.351  & -26.811  & \multicolumn{1}{c|}{-40.211 } & 7.982  \\
    \midrule
    \multicolumn{1}{c|}{\multirow{3}[2]{*}{EMO-NAS-CD}} & S1    & -0.375  & 0.469  & -0.762  & \multicolumn{1}{c|}{0.832 } & -1.148  & -1.394  & -3.878  & \multicolumn{1}{c|}{-5.167 } & 0.023  & 0.258  & 0.012  & \multicolumn{1}{c|}{0.316 } & -9.525  & -17.692  & -26.303  & \multicolumn{1}{c|}{-38.852 } & 6.688  \\
    \multicolumn{1}{c|}{} & S2    & -0.954  & -1.897  & -0.836  & \multicolumn{1}{c|}{1.237 } & -0.990  & -1.449  & -2.816  & \multicolumn{1}{c|}{-4.500 } & 0.224  & -0.295  & -0.047  & \multicolumn{1}{c|}{0.624 } & -9.590  & -17.107  & -26.213  & \multicolumn{1}{c|}{-37.099 } & 6.617  \\
    \multicolumn{1}{c|}{} & S3    & -1.137  & -2.024  & -6.011  & \multicolumn{1}{c|}{3.462 } & -0.550  & -2.724  & -3.149  & \multicolumn{1}{c|}{-2.149 } & 1.900  & -0.812  & -1.462  & \multicolumn{1}{c|}{0.187 } & -8.623  & -16.471  & -23.519  & \multicolumn{1}{c|}{-34.791 } & 6.811  \\
    \midrule
    \multicolumn{1}{c|}{\multirow{5}[2]{*}{OSCD}} & SA1   & -0.371  & -0.568  & -0.452  & \multicolumn{1}{c|}{0.487 } & -0.997  & -1.414  & -2.295  & \multicolumn{1}{c|}{-4.405 } & -0.162  & -0.128  & 0.046  & \multicolumn{1}{c|}{-0.336 } & -7.408  & -15.650  & -23.974  & \multicolumn{1}{c|}{-41.387 } & 6.255  \\
    \multicolumn{1}{c|}{} & SA2   & -0.267  & -0.394  & -0.522  & \multicolumn{1}{c|}{0.603 } & -1.021  & -1.427  & -2.610  & \multicolumn{1}{c|}{-4.291 } & -0.290  & -0.058  & -0.035  & \multicolumn{1}{c|}{-0.232 } & -7.469  & -15.623  & -23.788  & \multicolumn{1}{c|}{-35.212 } & 5.865  \\
    \multicolumn{1}{c|}{} & SA3   & -0.302  & -0.487  & -0.337  & \multicolumn{1}{c|}{0.789 } & -0.940  & -1.555  & -2.518  & \multicolumn{1}{c|}{-4.201 } & -0.104  & -0.255  & 0.046  & \multicolumn{1}{c|}{-0.197 } & -7.264  & -15.665  & -24.054  & \multicolumn{1}{c|}{-35.113 } & 5.864  \\
    \multicolumn{1}{c|}{} & SA4   & -0.139  & -0.407  & -0.570  & \multicolumn{1}{c|}{0.825 } & -0.756  & -1.360  & -2.511  & \multicolumn{1}{c|}{-3.975 } & -0.012  & -0.151  & -0.070  & \multicolumn{1}{c|}{0.244 } & -7.288  & -16.773  & -23.794  & \multicolumn{1}{c|}{-34.848 } & 5.858  \\
    \multicolumn{1}{c|}{} & SA5   & -0.186  & -0.221  & -0.465  & \multicolumn{1}{c|}{0.569 } & -0.732  & -1.220  & -2.393  & \multicolumn{1}{c|}{-4.077 } & 0.000  & 0.058  & -0.070  & \multicolumn{1}{c|}{0.012 } & -8.316  & -16.562  & -23.891  & \multicolumn{1}{c|}{-36.272 } & 5.940  \\
    \midrule
    \multicolumn{2}{c|}{+/-} & 0/13  & 3/10 & 1/12 & 2/11 & 1/12 & 1/12 & 0/13  & 1/12 & 3/10 & 3/10 & 1/12 & 1/12 & 1/12 & 1/12 & 2/11 & 3/10 & 0/13 \\
    \bottomrule
    \end{tabular}%
}
\end{table*}%

\subsection{Searching Stage}
\subsubsection{Individual Representation and Population Initialization}
Following prior works, we first define individual representation. To precisely encode candidate architectures from the search space~(Figure~\ref{fig:search_space1}) into evolvable individuals, we utilize a fixed-length integer vector $\mathbf{Individual}_i=[ind_1,\dots,ind_{2^{h-1}},ind_{2^{h-1}+1},\dots,ind_{2^h-1}]$
where $h$ is the height of the full binary tree topology. Each element in $\mathbf{Individual}_i$ corresponds to a configurable decision point within the search space, such as input selection or operator choice. The integer value at each position explicitly specifies the concrete configuration option. Most crucially, we design a tailored population initialization mechanism. When constructing the initial population of size $Pop$, we concurrently deploy: uniform random sampling for comprehensive search space coverage, and injection of elite genes from validated cognitive diagnosis models such as IRT, MIRT, and NCD. This dual-strategy framework leverages prior knowledge to accelerate convergence while maintaining topological diversity, ultimately enhancing Pareto frontier approximation efficiency without compromising global search capacity.

\subsubsection{Objective Function}
To simultaneously optimize candidate architectures for accuracy on original validation data and robustness in complex heterogeneous noise scenarios while ensuring the distribution characteristics of model predictions under noise interference closely match those on original data, we define the following three objective functions for joint optimization to enhance behavioral consistency and stability at the output level:
\begin{equation}
    \max_{Archi}F(Archi)=\left\{
        \begin{aligned}
         f_1(Archi) & =AUC(Archi; \mathcal{R}_\mathrm{Val}) \\
         f_2(Archi) & =AUC(Archi; \tilde{\mathcal{R}}_\mathrm{Val}) \\
         f_3(Archi) & =D_{\mathrm{KL}}(\mathcal{P} || \tilde{\mathcal{P}};Archi)
        \end{aligned}
        \right.,
\end{equation}
where $Archi$ represents the candidate architecture to be optimized, $AUC(Archi;D)$ represents the AUC~(Area Under an ROC Curve) of $Archi$ on the dataset $D$, $\mathcal{P}$ and $\tilde{\mathcal{P}}$ represents the predicted probability distributions of $archi$ on $\mathcal{R}_\mathrm{Val}$ and $\tilde{\mathcal{R}}_\mathrm{Val}$ respectively, and $D_{\mathrm{KL}}$ is the Kullback-Leibler divergence between these two probability distributions. This objective function drives the search process to simultaneously optimize three critical dimensions:

(1) \textbf{Fundamental Performance Guarantee:} $f_1$ uses AUC as the performance baseline to ensure the model possesses essential discriminative characteristics in clean data domains. This dimension directly reflects the expected prediction accuracy of the model in ideal environments.

(2) \textbf{Noise Robustness Characterization:} $f_2$ systematically quantifies operational fluctuations in non-ideal environments by constructing heterogeneous noise scenarios $\tilde{\mathcal{R}}_\mathrm{Val}=\{\mathcal{R}_\mathrm{Val}^\Theta, \mathcal{R}_\mathrm{Val}^\Phi, \mathcal{R}_\mathrm{Val}^\Psi, \\ \mathcal{R}_\mathrm{Val}^\Omega \}$ (including multi-source corruptions like Log Miss, Exercise Confusion, Q-matrix Confusion, and Log Flip). This reflects the model's performance instability when facing heterogeneous perturbations.

(3) \textbf{Distribution Consistency Constraint:} $f_3$ employs KL divergence minimization as the distribution alignment criterion, enforcing functional convergence between clean data prediction perturbation $\mathcal{P}$ and noisy environment perturbation $\tilde{\mathcal{P}}$. This mechanism explicitly penalizes prediction behavior deviations, fundamentally ensuring the topological invariance of model behavior under perturbations.
Specifically, by minimizing the KL divergence between distributions, we effectively encourage the model to maintain consistency in its output distribution when exposed to perturbations or noise. This mechanism prevents the model from overfitting to random fluctuations in individual samples, thereby mitigating the risk of reduced robustness.
The proof is provided in \textbf{Appendix A.6}.

\subsubsection{Genetic Operations}
For fixed-length integer vectors (length $2^h-1$) representing full binary trees of height h, we employ the standard single-point crossover and the bit-wise mutation~\cite{SparseEA} operations. The single-point crossover randomly selects a position in the vector and exchanges the gene segments before and after this position between two parent individuals to produce offspring. Then, the bit-wise mutation randomly selects a single gene locus in the vector and modifies its value to another valid value within the defined domain. These operations, performed within the closed integer vector space $\mathbb{Z}^{2^h-1}$, maintain an isomorphic mapping to the tree structure through the encoding scheme, strictly preserving the topological integrity of the full binary tree.

\begin{table*}[t]
\centering
\caption{Runtime comparison of CDMs on ASSIST09 and SLP-Math.}
\vspace{-3mm}
\label{tab:time_cost}
\resizebox{0.98\linewidth}{!}{
    \begin{tabular}{c|c|ccccccccccc|c}
    \toprule
    \multirow{3}[4]{*}{Time Cost(minutes)} & Dataset & DINA  & IRT   & MIRT  & MF    & NCD   & RCD   & KSCD  & KaNCD & ReliCD & ORCDF & EMO-NAS-CD & OSCD \\
\cmidrule{2-14}          & ASSIST09 & 9.2   & 17.8  & 24.2  & 13.1  & 18.4  & 486.8 & 15.9  & 13.8  & 94.1  & 37.2  & 15.2 GPU days & 26.1 hours \\
          & SLP-Math & 0.5   & 1.1   & 0.8   & 0.9   & 1.8   & 31.3  & 0.7   & 1.4   & 48.6  & 5.5   & 30.6 hours & 51.7 mins \\
    \bottomrule
    \end{tabular}%
}
\end{table*}%

\section{EXPERIMENT}
In this section, we conduct comprehensive experiments on two publicly available real-world educational datasets to validate the effectiveness of our proposed approach. Specifically, we will answer the following research questions (\textbf{RQs}) to unfold the experiments:
\begin{itemize}[itemsep=2pt,topsep=0pt,parsep=0pt, leftmargin=15pt]
    \item $\textbf{RQ1}$: How does the proposed OSCD model demonstrate effectiveness and superiority in addressing the CD task?
    \item $\textbf{RQ2}$: How does the robustness of OSCD compare to other baseline methods under different types of noise?
    \item $\textbf{RQ3}$: Can OSCD's supernet reliably predict the relative performance of random architectures?
    \item $\textbf{RQ4}$: Is OSCD efficient in terms of architecture search time?
    \item $\textbf{RQ5}$: What insights can be gained from visualizing OSCD’s searched architectures and their objective function values?
\end{itemize}

\subsection{Experimental Setting}
\subsubsection{\textbf{Datasets.}}
To assess the effectiveness of the proposed model in learner modeling, we conducted experiments on two publicly available real-world educational datasets: ASSIST09\cite{Assistment0912}, SLP-Math\cite{SLP}. More details of the dataset descriptions are available in \textbf{Appendix A.8}.

\subsubsection{\textbf{Baseline Methods.}}
We conducted extensive experiments and compared several strong and commonly used baselines, including DINA\cite{DINA}, IRT\cite{IRT}, MIRT\cite{MIRT}, MF\cite{MF}, NCD\cite{NCD}, RCD\cite{RCD}, KSCD\cite{KSCD}, KaNCD\cite{NeuralCD}, ReliCD\cite{ReliCD}, ORCDF\cite{ORCDF}, and EMO-NAS-CD\cite{NASCD}. The introduction and implementation details can be found in \textbf{Appendix A.9}. We evaluated the models utilizing three common metrics, including  ACC, AUC, and RMSE.


\subsubsection{\textbf{Implementation Details.}}
Following the approach described in \cite{NCD,CTNC}, for each dataset, we filtered out learners with fewer than $15$ historical response records to ensure that every learner has sufficient exercise records for diagnosis. All experimental results are reported as averages computed over ten independent runs. We implemented all methods with PyTorch by Python, and the experiments were conducted on an NVIDIA Geforce RTX4090 GPU.
For the baselines, we utilized the optimal parameter configurations suggested in the original work. In training stage, the embedding size $d$, batch size, and learning rate were set to $128$, $128$, $0.001$, respectively. The height $h$ of tree was set to $6$. In searching stage, the population size $Pop$ was set to $100$, the maximal number of generation was set to $100$. Our codes are available at https://github.com/AhuTwelve/OSCD.

\begin{figure}[t]
\centering
\includegraphics[width=\linewidth]{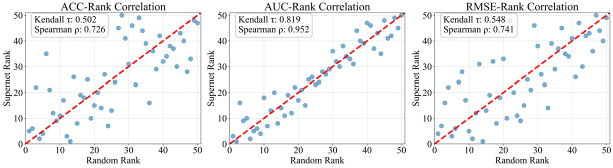}
\vspace{-3mm}
\caption[t]{Ranking fidelity evaluation between supernet-based predictions and independent training results.}
\label{fig:supernet_rank}
\end{figure}

\subsection{Performance Comparison}
Table~\ref{tab:main_experiment} compares the ACC, AUC, and RMSE of the baselines, the top-performing architectures discovered by EMO-NAS-CD, and those discovered by our proposed OSCD on two educational datasets: ASSIST09 and SLP-Math. The best results for each model are highlighted in bold, while the second-best results are underlined. Table~\ref{tab:main_experiment}(a) and Table~\ref{tab:main_experiment}(b) respectively present the performance comparisons of various methods under different datasets and their corresponding split ratios, aiming to validate the robustness and effectiveness of our method across diverse experimental settings. 
According to the results, there are several observations: (1) The architectures discovered by our proposed OSCD method achieved the best or second-best performance on both datasets, significantly outperforming most existing baseline models overall, which fully demonstrates the effectiveness of the proposed method. Notably, on the SLP-Math dataset, the discovered architectures SA1 and SA5 achieved the best results across almost all evaluation metrics, further validating OSCD's capability in architecture optimization when dealing with real-world educational data.
(2) Under various split settings, the discovered architectures maintained stable performance across different metrics, showing minimal fluctuations and demonstrating strong generalization and resistance to overfitting. Compared with certain baselines such as IRT, MF, and KaNCD, which suffer from performance degradation under limited training data, the robustness of the models discovered by OSCD is particularly noteworthy.

\subsection{Robustness Study}
To verify the robustness of the model architectures obtained by OSCD, we simulate different heterogeneous noise environments on the original test set and inject noise at varying ratios. By comparing the AUC change ratios under different noise ratios, we can effectively assess the model’s stability under noise interference. A positive change ratio indicates that the model’s AUC improves under noise interference, while a negative change ratio indicates a decline in AUC. The absolute value of the change ratio reflects the magnitude of performance fluctuations under noise; the smaller the absolute value, the less sensitive the model is to noise, and hence, the stronger its robustness. 
As illustrated in Table~\ref{tab:robustness_study}, there are several noteworthy observations: (1) Across all noise types and ratio settings, the architectures SA1 to SA5 discovered by OSCD consistently exhibit the smallest average AUC change ratios, indicating the lowest susceptibility to noise and the smallest performance fluctuations. Compared with other baseline methods, OSCD achieves significantly lower fluctuation magnitudes in most cases, demonstrating stronger robustness and stability.
(2) Regarding the impact of heterogeneous noise types, log miss and $Q$-matrix confusion have relatively minor interference with model performance, whereas the other two types of noise cause severe damage to the models, particularly log flip, which can degrade the predictive performance of most models to a level close to random guessing.

\begin{figure}
    \centering
    \includegraphics[width=\linewidth]{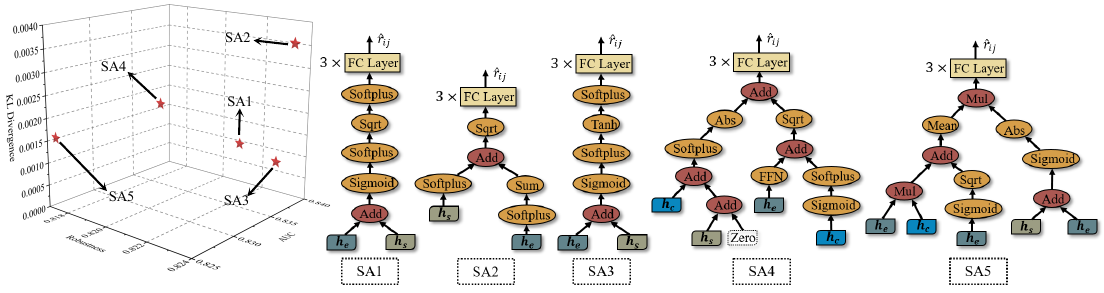}
    \vspace{-3mm}
    \caption{Visualization of Non-dominated individuals on the SLP dataset.}
    \label{fig:vis_slp}
\end{figure}

\begin{figure}
    \centering
    \includegraphics[width=\linewidth]{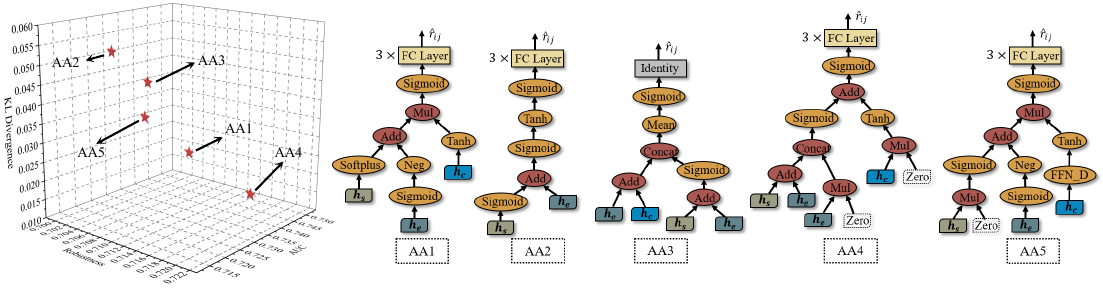}
    \vspace{-3mm}
    \caption{Visualization of Non-dominated individuals on the ASSIST09 dataset.}
    \label{fig:vis_ASSIST09}
\end{figure}

\subsection{Ranking Correlation Evaluation}
To evaluate the fidelity of the supernet’s performance prediction, we randomly generated 50 architectures and compared the rankings obtained from direct supernet evaluation with those from independently trained models, calculating Kendall’s tau and Spearman’s rank correlation coefficients to quantify the consistency between the two evaluation results. As depicted in the Figure~\ref{fig:supernet_rank}, it can be observed that the performance rankings predicted by the supernet are highly consistent with those obtained from independent training across the ACC, AUC, and RMSE evaluation metrics, with most data points lying along the diagonal, indicating strong ranking consistency. Moreover, the Kendall’s tau and Spearman’s rank correlation coefficients of AUC are close to $1$, further confirming the high fidelity of the supernet in performance prediction. This demonstrates that OSCD’s supernet can reliably reflect the true relative performance of random architectures, providing a solid foundation for efficient and accurate architecture search.

\subsection{Time Efficiency Analysis}
Table~\ref{tab:time_cost} presents the runtime comparison on ASSIST09 and SLP-Math. While traditional CDMs run quickly, their limited expressiveness constrains performance. EMO-NAS-CD achieves strong results but demands $15.2$ GPU days on ASSIST09 and $30.6$ hours on SLP-Math, making large-scale exploration impractical.
OSCD delivers a decisive efficiency breakthrough. It reduces runtime to $26.1$ hours on ASSIST09 ($14×$ faster) and $51.7$ minutes on SLP-Math ($35×$ faster), enabling rapid architecture exploration within feasible time budgets. This efficiency transforms NAS in cognitive diagnosis from a costly niche method into a practical, scalable tool, unlocking more iterative experimentation and broader research opportunities. By lowering the computational barrier, OSCD paves the way for more agile, exploratory, and impactful research in robust learner modeling and intelligent tutoring systems.

\subsection{Visualization of Discovered Architectures}
To further investigate the reasons behind the robustness of the searched architectures, we conducted additional visualization experiments. As shown in the Figure~\ref{fig:vis_slp} and \ref{fig:vis_ASSIST09}, the visualization results reveal that the non-dominated individuals generally select operations that satisfy the Lipschitz condition. Such operations can effectively constrain the sensitivity of the network outputs to input perturbations, which is a key factor in enhancing robustness. This phenomenon offers a new perspective for understanding the formation mechanism of robust architectures and may provide further inspiration for the design of future robust neural networks.

\section{CONCLUSION}
In this work, we proposed \textbf{OSCD}, an evolutionary multi-objective \underline{\textbf{O}}ne-\underline{\textbf{S}}hot neural architecture search framework tailored for \underline{\textbf{C}}ognitive \underline{\textbf{D}}iagnosis, aiming to identify both efficient and robust model architectures. Specifically, OSCD proceeds in two stages: a training stage and a searching stage. During training, a comprehensive search space covering diverse architectural configurations is established, and a weight-sharing supernet based on a complete binary tree topology is optimized, enabling the exploration of architectures beyond hand-crafted design constraints. In the subsequent searching stage, the problem of identifying the optimal architecture under heterogeneous noise conditions is formulated as a multi-objective optimization task. This is addressed through a framework that combines Pareto-optimal solution discovery with cross-scenario performance evaluation. Extensive experiments on real-world educational datasets demonstrate that the architectures obtained through OSCD consistently deliver effective and robust performance in CD tasks, underscoring the practical value of the proposed approach. We hope this work will provide new perspectives in the field of CD.

\begin{acks}
This work was supported in part by the National Natural Science Foundation of China under Grant (NO. 62577002 and No.62302010), in part by the Anhui Provincial Natural Science Foundation (NO. 2508085MF160), in part by China Postdoctoral Science Foundation (No.2023M740015), in part by the Postdoctoral Fellowship Program (Grade B) of China Postdoctoral Science Foundation (No.GZB20240002), and in part by the Anhui Province Key Laboratory of Intelligent Computing and Applications (No. AFZNJS2024KF01).
\end{acks}

\clearpage
\bibliographystyle{ACM-Reference-Format}
\balance
\bibliography{ref}

\appendix

\section{Appendix}

\subsection{RELATED WORK}
\subsubsection{Cognitive Diagnosis}
Cognitive diagnosis, as an advanced learner modeling approach, is theoretically grounded in the intersection of modern psychometrics and cognitive science. By analyzing students' interactive learning behaviors, it can systematically deconstruct learners' knowledge states and cognitive processes.
Current research primarily focuses on manually developing sophisticated cognitive diagnostic models (CDMs) for more accurate learner modeling and assessment, which can be broadly classified into three categories: psychometric theory-based methods, neural network-based methods, and graph modeling-based methods. In the early stages of cognitive diagnosis development, researchers mainly employed psychometric techniques to assess learners' abilities, utilizing unidimensional or multidimensional latent vectors to measure skill proficiency, as exemplified by DINA\cite{DINA}, IRT\cite{IRT}, and MIRT\cite{MIRT}. 
The advancement of deep learning has sparked significant interest, leading to the emergence of various neural network-based approaches~\cite{NCD,KSCD,DZCD}. For instance, NCD\cite{NCD}, as a representative neural CDM, employs multidimensional parameters to construct fine-grained representations of students' cognitive states and exercise attributes, while leveraging neural networks to model complex interactions. KSCD\cite{KSCD} feeds student, exercise, and concept vectors into an IRT-like neural network as a predictive diagnostic function to uncover implicit relationships among concepts. Similarly, domain-level zero-shot cognitive diagnosis\cite{DZCD} and diagnosis using incremental response logs\cite{ICD} have been developed to address practical challenges in online learning environments.

Owing to the unique characteristics of graph structures and the advancements in graph learning, graph modeling-based methods\cite{RCD,DGCD,ORCDF,RDGT} have gained widespread acceptance. These methods primarily leverage graph topological information to reveal higher-order relationships among students, exercises, and concepts, as well as to mine valuable latent implicit associations. For example, RCD\cite{RCD} models student interactions and structural relationships by constructing a student-exercise-concept graph, employing graph convolutional networks (GCN\cite{GCN}) to enhance the effectiveness of representation learning. With the increasing attention to learner modeling, recent years have also witnessed critical explorations from other perspectives, including interpretability\cite{ID-CDF,KanCD}, robustness\cite{AdaRD}, and group learning\cite{HomoGCD,RIGL}.

\subsubsection{Neural Architecture Search}
As a revolutionary automated machine learning technology\cite{NASNet,NASsurvey2}, Neural Architecture Search (NAS) has evolved into a comprehensive multi-domain and multi-modal framework since its pioneering works\cite{NAS,NASsurvey1,ENASsurvey1}. Currently, NAS techniques have achieved cross-domain penetration by offering automated optimization solutions for a wide range of mainstream deep neural networks (DNNs). This technology has not only been successfully applied to the construction of core models like CNNs in computer vision (CV)\cite{BigNAS}, but has also extended to recurrent neural networks (RNNs) in natural language processing (NLP)\cite{TextNAS}, speech recognition systems\cite{ASRNAS}, and graph neural networks (GNNs) for tasks such as node classification and link prediction\cite{GraphNASsurvey1}. Notably, the Transformer architecture, as a universal paradigm for cross-modal applications, has also benefited from the systematic exploration enabled by NAS methods\cite{ViTAS,TransformerNASsurvey1,NAS-BERT,Darts-Conformer}.

\subsection{Pseudocode of OSCD}
Next, we present the core algorithm pseudocode of OSCD to facilitate a thorough understanding of its internal working mechanism.
\begin{algorithm}
    \caption{Main Steps of OSCD} 
    \label{algo} 
    \begin{algorithmic}[1]
        \STATE \textbf{Input:} $Gen$: Maximum number of generations; $Pop$: Population size; $S$: Search space
        \STATE \textbf{Output:} $\mathbf{P}_b$: The best individual;
        \STATE $\mathbf{P}$ $\leftarrow$ Randomly initialize $Pop$ individuals;
        \STATE Evaluate each individual in $\mathbf{P}$ based on the trained supernet and obtain its AUC value;
        \FOR {$g = 1$ to $Gen$}
        \STATE $\mathbf{P}'$ $\leftarrow$ Select $Pop$ parent individuals from Population $\mathbf{P}$;
        \STATE $\mathbf{Q}$ $\leftarrow$ Apply the single-point crossover and bit-wise mutation to $\mathbf{P}'$ based on current search space $S$;
        \STATE Evaluate each individual in $\mathbf{Q}$ based on the trained supernet and obtain the objective values;
        \STATE $\mathbf{P}$ $\leftarrow$ Environment Selection($\mathbf{P}\cup \mathbf{Q}$);
        \ENDFOR
        \STATE Select the best individual $\mathbf{P}_b$ from $\mathbf{P}$ as the output;
        \STATE \textbf{return} $\mathbf{P_b}$.
    \end{algorithmic} 
\end{algorithm}

\subsection{Proof of Lipschitz Condition}
We provide more detailed supplementary explanations for the operators in the table.
First, it is worth noting that $FFN$, $FFN\_D$ and $Concat$ are all FC-layer-based operations, indicating they incorporate learnable parameters to enhance the representational capacity of the model architecture. Moreover, while other operators contain no learnable parameters, the embedding layer in Eq.\eqref{eq:1} ensures the general model avoids being limited to simple symbolic computations. Next, we introduce the definition and proof of the Lipschitz condition.

\textbf{Definition 4.1(Lipschitz Condition).} \textit{Let $f: D\rightarrow \mathbb{R}$ be a function defined on a set $D\subseteq \mathbb{R}$. The function $f$ is said to satisfy that Lipschitz condition (or be Lipschitz continuous) on $D$ if there exists a constant $L \geq 0$ such that}
\begin{equation}
    |f(x)-f(y)| \leq L|x-y|, \forall x,y\in D.
\end{equation}
\textit{Any constant $L$ satisfying this inequality is called a Lipschitz constant for $f$ on $D$. The Lipschitz constant of $f$, denoted $L_f$, is the infimum (greatest lower bound) of all Lipschitz constants for $f$ on $D$; that is,}
\begin{equation}
    L_f:=inf\{L \geq 0 ||f(x)-f(y) \leq L|x-y|,\forall x,y\in D\}.
\end{equation}

\textit{Proof:}

\begin{itemize}[leftmargin=10pt, itemsep=2pt, topsep=0pt, parsep=0pt]
    \item \textbf{Iden(\ding{52}):} Given that $x$, $y$ and $Iden(x)=x$, we then have
    \begin{equation}
        |Iden(x)-Iden(y)|=|x-y|\leq 1\cdot |x-y|.
    \end{equation}
    Therefore, it can be proved that $Iden$ operator satisfies the Lipschitz condition, and the Lipschitz constant is $L=1$.
    \item \textbf{Neg(\ding{52}):} Given that $x$, $y$ and $Neg(x)=-x$, we then have
    \begin{equation}
        |Neg(x)-Neg(y)|=|y-x|=|x-y|\leq 1\cdot |x-y|.
    \end{equation}
    Therefore, it can be proved that $Neg$ operator satisfies the Lipschitz condition, and the Lipschitz constant is $L=1$.
    \item \textbf{Abs(\ding{52}):} Given that $x$, $y$ and $Abs(x)=|x|$, we then have
    \begin{equation}
        |Abs(x)-Abs(y)|=||x|-|y||\leq |x-y|= 1\cdot |x-y|.
    \end{equation}
    Therefore, it can be proved that $Abs$ operator satisfies the Lipschitz condition, and the Lipschitz constant is $L=1$.
    \item \textbf{Inv(\ding{56}):} Given that $x$, and $Inv(x)=1/(x+1e^{-6})$, we can obtain
    \begin{equation}
        Inv'(x)=-\frac{1}{(x+1e^{-6})^2},
    \end{equation}
    when $x\rightarrow -1e^{-6}$, $|Inv'(x)|\rightarrow \infty$. Based on the definition of Lipschitz condition, if the derivative of a function is unbounded, then the function does not satisfy the Lipschitz condition. Therefore, it can be proved that $Inv$ operator does not satisfy the Lipschitz condition.
    \item \textbf{Square(\ding{56}):} Given that $x$, and $Square(x)=x^2$, we can obtain
    \begin{equation}
        Square'(x)=2x,
    \end{equation}
    when $x\rightarrow \infty$, $|Square'(x)|\rightarrow \infty$. According to the definition of Lipschitz condition, if the derivative of a function is unbounded, then the function does not satisfy the Lipschitz condition. Therefore, it can be proved that $Square$ operator does not satisfy the Lipschitz condition.
    \item \textbf{Sqrt(\ding{56}):} Given that $x$, and $Sqrt(x)=sign(x)\cdot \sqrt{|x|+1e^{-6}}$, we can obtain
    \begin{equation}
        Sqrt'(x)=\left\{
        \begin{aligned}
         \frac{1}{2\sqrt{x+1e^{-6}}}, x\geq 0 \\
         \frac{1}{2\sqrt{1e^{-6}-x}}, x<0
        \end{aligned}
        \right.,
    \end{equation}
    when $x\rightarrow 0$, $Sqrt'(x)\rightarrow 1/{2\sqrt{1e^{-6}}}$, where $1e-6$ is the special value we set. Theoretically, $Sqrt(x)$ is not continuous at $x=0$ in terms of its derivative. Therefore, we believe that $Sqrt$ operator does not satisfy the Lipschitz condition.
    \item \textbf{Tanh(\ding{52}):} Given that $x$, $y$ and $Tanh(x)=tanh(x)=(e^x-e^{-x})/(e^x+e^{-x})$, we can obtain
    \begin{equation}
        tanh'(x)=1-tanh^2(x),
    \end{equation}
    because of $tanh(x)\in(-1,1)$, we can obtain that $tanh'(x)\in(0,1]$, and
    \begin{equation}
        |Tanh(x)-Tanh(y)|=|tanh(x)-tanh(y)|\leq 1\cdot |x-y|.
    \end{equation}
    Therefore, it can be proved that $Tanh$ operator satisfies the Lipschitz condition, and the Lipschitz constant is $L=1$.
    \item \textbf{Sigmoid(\ding{52}):} Given that $x$, $y$ and $Sigmoid(x)=sigmoid(x)=1/(1+e^{-x})$, we can obtain
    \begin{equation}
        sigmoid'(x)=sigmoid(x)\cdot (1-sigmoid(x)),
    \end{equation}
    because of $sigmoid(x)\in(0,1)$, we can obtain that $sigmoid'(x)\in(0,\frac{1}{4}]$, and
    \begin{equation}
        |Sigmoid(x)-Sigmoid(y)|=|sigmoid(x)-sigmoid(y)|\leq \frac{1}{4}\cdot |x-y|.
    \end{equation}
    Therefore, it can be proved that $Sigmoid$ operator satisfies the Lipschitz condition, and the Lipschitz constant is $L=\frac{1}{4}$.
    \item \textbf{Softplus(\ding{52}):} Given that $x$, $y$ and $Softplus(x)=softplus(x)=\log(1+e^x)$, we can obtain
    \begin{equation}
        softplus'(x)=\frac{e^x}{1+e^x}=\frac{1}{1+e^{-x}}=sigmoid(x),
    \end{equation}
    because of $sigmoid(x)\in(0,1)$, we can obtain that $softplus'(x)\in(0,1)$, and
    \begin{equation}
        |softplus(x)-softplus(y)|=|softplus(x)-softplus(y)|\leq 1\cdot |x-y|.
    \end{equation}
    Therefore, it can be proved that $Softplus$ operator satisfies the Lipschitz condition, and the Lipschitz constant is $L=1$.
    \item \textbf{Sum(\ding{52}):} Given that $\mathbf{x}=(x_1, x_2, \dots, x_d)$, $\mathbf{y}=(y_1, y_2, \dots, y_d)$ and $Sum(\mathbf{x})=\sum_{i=1}^d x_i$, we then have
    \begin{equation}
        |Sum(\mathbf{x})-Sum(\mathbf{y})|=|\sum_{i=1}^d(x_i-y_i)|\leq \sum_{i=1}^d|x_i-y_i|,
    \end{equation}
    Based on the Cauchy-Schwarz inequality, for $\mathbf{a}=(a_1, a_2, \dots, a_d)$ and $\mathbf{b}=(b_1, b_2, \dots, b_d)$, we have
    \begin{equation}
        \sum_{i=1}^d|a_ib_i|\leq {||a||}_2 {||b||}_2,
    \end{equation}
    we take $\mathbf{a}=(1, 1, \dots, 1)$, $\mathbf{b}=(|x_1-y_1|, |x_2-y_2|, \dots, |x_d-y_d|)$ and we then obtain
    \begin{align}
    \nonumber
        \sum_{i=1}^d|x_i-y_i| & =\sum_{i=1}^d|1\cdot (x_i-y_i)|=\sum_{i=1}^d|a_ib_i|\\ & \leq {||a||}_2 {||b||}_2=\sqrt{d}\cdot {||x-y||}_2.
    \end{align}
    Therefore, it can be proved that $Sum$ operator satisfies the Lipschitz condition, and the Lipschitz constant is $L=\sqrt{d}$.
    \item \textbf{Mean(\ding{52}):} Given that $\mathbf{x}=(x_1, x_2, \dots, x_d)$, $\mathbf{y}=(y_1, y_2, \dots, y_d)$ and $Mean(\mathbf{x})=\sum_{i=1}^d x_i/d$, we then have
    \begin{equation}
        |Mean(\mathbf{x})-Mean(\mathbf{y})|=|\frac{1}{d}\sum_{i=1}^d(x_i-y_i)|\leq \frac{1}{d}\sum_{i=1}^d|x_i-y_i|,
    \end{equation}
    Based on the Cauchy-Schwarz inequality, for $\mathbf{a}=(a_1, a_2, \dots, a_d)$ and $\mathbf{b}=(b_1, b_2, \dots, b_d)$, we have
    \begin{equation}
        \sum_{i=1}^d|a_ib_i|\leq {||a||}_2 {||b||}_2,
    \end{equation}
    we take $\mathbf{a}=(1, 1, \dots, 1)$, $\mathbf{b}=(|x_1-y_1|, |x_2-y_2|, \dots, |x_d-y_d|)$ and we then obtain
    \begin{align}
    \nonumber
        \frac{1}{d}\sum_{i=1}^d|x_i-y_i| & =\frac{1}{d}\sum_{i=1}^d|1\cdot (x_i-y_i)|=\frac{1}{d}\sum_{i=1}^d|a_ib_i| \\ & \leq \frac{1}{d}{||a||}_2 {||b||}_2=\frac{1}{\sqrt{d}}\cdot {||x-y||}_2.
    \end{align}
    Therefore, it can be proved that $Mean$ operator satisfies the Lipschitz condition, and the Lipschitz constant is $L=1/\sqrt{d}$.
    \item \textbf{FFN(\ding{52}):} Given that $\mathbf{x}$, $\mathbf{y}$, and $FFN(\mathbf{x}) = \mathbf{x}\times \mathrm{W}_{ffn}$, we then have
    \begin{equation}
        ||FFN(\mathbf{x})-FFN(\mathbf{y})||=||(\mathbf{x}-\mathbf{y})\cdot \mathrm{W}_{ffn}|| \leq ||\mathbf{x}-\mathbf{y}||\cdot ||\mathrm{W}_{ffn}||_2.
    \end{equation}
    Therefore, it can be proved that $FFN$ operator satisfies the Lipschitz condition, and the Lipschitz constant is $L=||\mathrm{W}_{ffn}||_2$.
    \item \textbf{FFN\_D(\ding{52}):} Given that $\mathbf{x}$, $\mathbf{y}$, and $FFN\_D(\mathbf{x})=\mathbf{x}\times \mathrm{W}_{ffnd}$, we then have 
    \begin{equation}
        ||FFN\_D(\mathbf{x})-FFN\_D(\mathbf{y})||=||(\mathbf{x}-\mathbf{y})\cdot \mathrm{W}_{ffnd}|| \leq ||\mathbf{x}-\mathbf{y}||\cdot ||\mathrm{W}_{ffnd}||_2.
    \end{equation}
    Therefore, it can be proved that $FFN\_D$ operator satisfies the Lipschitz condition, and the Lipschitz constant is $L=||\mathrm{W}_{ffnd}||_2$.
    \item \textbf{Add(\ding{52}):} Given that $\mathbf{x}=(x_1,x_2,\dots,x_d)$, $\mathbf{y}=(y_1,y_2,\dots,y_d)$, $\mathbf{x'}=(x'_1,x'_2,\dots,x'_d)$, $\mathbf{y'}=(y'_1,y'_2,\dots,y'_d)$ and $Add(\mathbf{x},\mathbf{y})=\mathbf{x}+\mathbf{y}$, we then have
    \begin{align}
        \nonumber
        & ||Add(\mathbf{x},\mathbf{y})-Add(\mathbf{x}',\mathbf{y}')|| \\ 
        & =||\mathbf{x}-\mathbf{x}'+\mathbf{y}-\mathbf{y}'|| \\ 
        \nonumber & \leq ||\mathbf{x}-\mathbf{x}'|| + ||\mathbf{y}-\mathbf{y}'||.
    \end{align}
    Therefore, it can be proved that $Add$ operator satisfies the Lipschitz condition, and the Lipschitz constant is $L=1$.
    \item \textbf{Mul(\ding{56}):} Given that $\mathbf{x}=(x_1,x_2,\dots,x_d)$, $\mathbf{y}=(y_1,y_2,\dots,y_d)$, $\mathbf{x'}=(x_1+\delta,x_2+\delta,\dots,x_d+\delta)$, $\mathbf{y'}=(y_1+\delta,y_2+\delta,\dots,y_d+\delta)$ and $Mul(\mathbf{x},\mathbf{y})=\mathbf{x}\odot \mathbf{y}$, we then have
    \begin{align}
        \nonumber
        & ||Mul(\mathbf{x}, \mathbf{y})-Mul(\mathbf{x}', \mathbf{y}')|| \\
        \nonumber & = ||\mathbf{x} \odot \mathbf{y}-\mathbf{x}'\odot \mathbf{y}'|| \\
        \nonumber & = ||(x_1y_1,\dots,x_dy_d)-((x_1+\delta)(y_1+\delta),\dots,(x_d+\delta)(y_d+\delta))|| \\
        & = ||(-(x_1+y_1)\delta-\delta^2),\dots,(-(x_d+y_d)\delta-\delta^2)|| \\
        \nonumber & = ||-\delta((x_1+y_1)+\delta),\dots,-\delta((x_d+y_d)+\delta)||,
    \end{align}
    because of $||\mathbf{x}-\mathbf{x}'||+||\mathbf{y}-\mathbf{y}'||=2\sqrt{d}\delta$, we can obtain
    \begin{align}
        \nonumber
        & \frac{||Mul(\mathbf{x}, \mathbf{y})-Mul(\mathbf{x}', \mathbf{y}')||}{||\mathbf{x}-\mathbf{x}'||+||\mathbf{y}-\mathbf{y}'||} \\
        & = \frac{||-\delta((x_1+y_1)+\delta),\dots,-\delta((x_d+y_d)+\delta)||}{2\sqrt{d}\delta} \\
        \nonumber & = \frac{||-((x_1+y_1)+\delta),\dots,-((x_d+y_d)+\delta)||}{2\sqrt{d}}.
    \end{align}
    Therefore, when $\mathbf{x},\mathbf{y}\rightarrow \infty$, $\frac{||Mul(\mathbf{x}, \mathbf{y})-Mul(\mathbf{x}', \mathbf{y}')||}{||\mathbf{x}-\mathbf{x}'||+||\mathbf{y}-\mathbf{y}'||}\rightarrow \infty$. It can be proved that $Mul$ operator does not satisfy the Lipschitz condition.
    \item \textbf{Concat(\ding{52}):} Given that $(\mathbf{x}_1,\mathbf{y}_1)$, $(\mathbf{x}_2,\mathbf{y}_2)$, and $Concat(\mathbf{x},\mathbf{y})=[\mathbf{x} || \mathbf{y}]\times \mathrm{W}_{concat}$, we then have
    \begin{align}
        \nonumber
        & ||Concat(\mathbf{x}_1, \mathbf{y}_1)-Concat(\mathbf{x}_2,\mathbf{y}_2)|| \\ \nonumber & =||([\mathbf{x}_1||\mathbf{y}_1]-[\mathbf{x}_2||\mathbf{y}_2]) \cdot \mathrm{W}_{concat}|| \\ & \leq ||([\mathbf{x}_1||\mathbf{y}_1]-[\mathbf{x}_2||\mathbf{y}_2]) \cdot ||\mathrm{W}_{concat}||_2 \\ \nonumber & \leq \sqrt{2} (||\mathbf{x}_1-\mathbf{x}_2||+||\mathbf{y}_1-\mathbf{y}_2||) \cdot ||\mathrm{W}_{concat}||_2.
    \end{align}
    Therefore, it can be proved that $Concat$ operator satisfies the Lipschitz condition, and the Lipschitz constant is $L=\sqrt{2} \cdot ||\mathrm{W}_{concat}||_2$.
\end{itemize}

\subsection{Challenges encountered during supernet training
}
We hereby provide a detailed mathematical formalization of the \textbf{Dimension Incompatibility} challenge. Furthermore, regarding our search space design and algorithm, we elucidate the advantages of all-ones padding.

(a) \textit{Unary operator mismatch}: Operators requiring single input are incorrectly connected to two input nodes. As illustrated in Figure~\ref{fig:search_space2}(a), given the triplet $node_1=\{input_1^L, input_1^R, operator_1\}$, where $input_1^L=\mathbf{h}_s\in \mathbb{R}^{1\times d}$, $input_1^R=\mathbf{h}_e\in \mathbb{R}^{1\times d}$, and $operator_1=Sigmoid(\cdot)$, since unary operators like $Sigmoid$ only accept a single input, we consistently feed the left input to the operator to ensure proper computation. That is to say, the computation process for the triplet $node_1$ is $sigmoid(\mathbf{h_s})$.

(b) \textit{Binary operator deficiency}: Operators requiring dual inputs are connected to only one input node. As shown in Figure~\ref{fig:search_space2}(b), given the triplet $node_1=\{input_1^L, input_1^R, operator_1\}$, where $input_1^L=\mathbf{h}_s\in \mathbb{R}^{1\times d}$, $input_1^R=Zero$, and $operator_1=Concat(\cdot~,\cdot)$, since $Zero$ merely serves as a placeholder without actual valid input, while $Concat$ requires two inputs, we adopt a dimension padding strategy: when the left node provides valid $d$-dimensional input, we generate a all-ones padding tensor of identical dimension for the right node. That is to say, the computation process of the triplet $node_1$ is $Concat(\mathbf{h}_s,\mathbf{1}^{1\times d})$.

(c) \textit{Dimension-sensitive operator left input error}: Dimension-sensitive operators receive a scalar as left input. As depicted in Figure~\ref{fig:search_space2}(c), the triplet $node_1=\{input_1^L, input_1^R, operator_1\}$, where $input_1^L=Sum(\mathbf{h}_s)\in \mathbb{R}^1$, $input_1^R=\mathbf{h}_e\in \mathbf{R}^{1\times d}$, and $operator_1=FFN(\cdot)$, has been provided. As previously mentioned, $FFN$ is a dimension-sensitive operator. When the left input dimension is not $d$-dimension, conventional padding methods would conflict with other binary operations. Therefore, we opt to replace the current operator with any non-dimension-sensitive operator to ensure training stability. For example, the computational process of this triplet $node_1$ then becomes $Sigmoid(Sum(\mathbf{h}_s))$.

(d) \textit{Dimension-sensitive operator right input error}: Dimension-sensitive operators receive $d$-dimension vectors as left input but the scalar as right input. As can be seen from Figure~\ref{fig:search_space2}(d), the left and right input dimensions of the triplet $node_1=\{input_1^L, input_1^R, operator_1\}$ are inconsistent, where $input_1^L= Sigmoid(\mathbf{h}_s)\in \mathbb{R}^{1\times d}$, $input_1^R=Sum(\mathbf{h}_e)\in \mathbb{R}^1$, $operator_1=Concat(\cdot~,\cdot)$. Given that the $operator_i$ already meets the scenario~(c), we employ an all-ones padding tensor to maintain the original computational path, thereby preventing the omission of critical operations through operator replacement. Other words, $Concat(Sigmoid(\mathbf{h}_s), [Sum(\mathbf{h}_e)|| \mathbf{1}^{1\times (d-1)}])$ denotes the computation process of the triplet $node_1$. 

It should be emphasized that we consistently employ all-ones padding tensors (rather than zero padding) to address dimension incompatibility issues. All-ones padding tensors minimize interference in computational processes within the defined search space, especially for operations like $Inv$, $Mul$, and various activation functions. This leads to better preservation of information flow integrity and gradient stability.

\subsection{Implementation details of the sandwich rule}
The Sandwich Rule is an efficient training strategy for weight-sharing NAS frameworks, designed to enhance the quality of supernet optimization. It addresses the performance estimation bias caused by inconsistent optimization difficulty across sub-models of varying sizes. The core methodology involves simultaneously training three types of subnets in each training iteration: the minimal subnet, the maximal subnet and randomly sampled subnets.
The minimal subnet, denoted as $\hat{r}_{ij}=Concat(\mathbf{h}_s,\mathbf{h}_e)$, typically represents the lower bound of the architectural space with the smallest parameter count and simplest computations in the search space. Similarly, the maximal subnet, denoted as $\hat{r}_{ij}=Concat(Concat(\dots \\(Concat(\mathbf{h}_s,\mathbf{h}_e))),Concat(\dots(Concat(\mathbf{h}_s,\mathbf{h}_c))))$, generally represents the upper bound of the architectural space with the largest parameter count and most complex computations in the search space. The randomly sampled subnets represent stochastic samples from the architectural space. 

\subsection{Theoretical Analysis of Objective Function}
\textbf{Theorem A.6}. \textit{Consider a model with predictive distributions $\mathcal{P}(y|x;\theta)$ on clean inputs $x$ and $\mathcal{P}(y|x+x';\theta)$ on noisy inputs, where the loss function is the cross-entropy loss. Assume that for all $x$, the two predictive distributions are sufficiently close. Then, there exists a constant $C>0$ such that the difference between the expected risks under clean and noisy conditions satisfies}
\begin{equation}
    |Risk_{noisy}-Risk_{clean}| \leq C||\mathcal{P}(y|x;\theta)-\mathcal{P}(y|x+x';\theta))||^2.
\end{equation}
    
\textit{Proof:}

For the cross-entropy loss function, we have

\begin{equation}
    \ell(Archi_\theta(x),y) = -\log \mathcal{P}(y|x;\theta).
\end{equation}

Define the KL divergence as

\begin{equation}
D_{KL}(\mathcal{P}(y|x;\theta)||\mathcal{P}(y|x+x');\theta)=\sum_y \mathcal{P}(y|x;\theta)\log \tfrac{\mathcal{P}(y|x;\theta)}{\mathcal{P}(y|x+x';\theta)},
\end{equation}

and the expected risks under clean and noisy conditions as

\begin{equation}
Risk_{clean}=\mathbb{E}_{x} \mathbb{E}_{y \sim \mathcal{P}(y|x)}[-\log \mathcal{P}(y|x;\theta)],
\end{equation}

\begin{equation}
Risk_{noisy}=\mathbb{E}_{x} \mathbb{E}_{y \sim \mathcal{P}(y|x+x')}[-\log \mathcal{P}(y|x+x';\theta)].
\end{equation}

Therefore, we obtain

\begin{align}
\nonumber & \Delta Risk = Risk_{noisy} - Risk_{clean} \\
\nonumber & =\mathbb{E}_{x} \mathbb{E}_{y \sim \mathcal{P}(y|x+x')}[-\log \mathcal{P}(y|x+x';\theta)] - \mathbb{E}_x \mathbb{E}_{y \sim \mathcal{P}(y|x)}[-\log \mathcal{P}(y|x;\theta)] \\
& =\mathbb{E}_x \sum_y [\mathcal{P}(y|x;\theta)\log \mathcal{P}(y|x;\theta) - \mathcal{P}(y|x+x';\theta)\log \mathcal{P}(y|x+x';\theta)].
\end{align}

Then, we introduce the intermediate term $\sum_y \mathcal{P}(y|x;\theta)\log \mathcal{P}(y|x+x';\theta)$ to decompose the expression 

\begin{align}
\nonumber \Delta Risk
\nonumber &= \mathbb{E}_x \sum_y [\mathcal{P}(y|x;\theta)\log \tfrac{\mathcal{P}(y|x;\theta)}{\mathcal{P}(y|x+x';\theta)} \\
\nonumber & \quad + (\mathcal{P}(y|x;\theta)-\mathcal{P}(y|x+x';\theta))\log \mathcal{P}(y|x+x';\theta)] \\
&= \mathbb{E}_x \sum_y [D_{KL}(\mathcal{P}(y|x;\theta)||\mathcal{P}(y|x+x');\theta) \\
\nonumber & \quad + (\mathcal{P}(y|x;\theta)-\mathcal{P}(y|x+x';\theta))\log \mathcal{P}(y|x+x';\theta)].
\end{align}

Next, to transform the above equation, we introduce another intermediate term $\sum_y \mathcal{P}(y|x+x';\theta)\log \mathcal{P}(y|x;\theta)$

\begin{align}
\nonumber \Delta Risk & = \mathbb{E}_x \sum_y [D_{KL}(\mathcal{P}(y|x;\theta)||\mathcal{P}(y|x+x');\theta) \\
& \quad - D_{KL}(\mathcal{P}(y|x+x';\theta)||\mathcal{P}(y|x);\theta) \\
\nonumber & \quad + \mathcal{P}(y|x;\theta)\log \mathcal{P}(y|x+x';\theta) - \mathcal{P}(y|x+x';\theta)\log \mathcal{P}(y|x;\theta)].
\end{align}

When the two distributions are close, i.e., $\mathcal{P}(y|x;\theta)\approx \mathcal{P}(y|x+x';\theta)$, we have

\begin{equation}
\mathcal{P}(y|x;\theta)\log \mathcal{P}(y|x+x';\theta) \approx \mathcal{P}(y|x+x';\theta)\log \mathcal{P}(y|x;\theta).
\end{equation}

Meanwhile, the difference in the bidirectional KL divergence is also a higher-order infinitesimal, $o(||\mathcal{P}(y|x;\theta)-\mathcal{P}(y|x+x';\theta))||^2)$, when the distributions are close. Therefore, there exists a constant $C>0$ such that

\begin{equation}
|\Delta Risk| \leq C||\mathcal{P}(y|x;\theta)-\mathcal{P}(y|x+x';\theta))||^2.
\end{equation}

As the predictive distribution $\mathcal{P}(y|x;\theta)$ becomes closer to the noisy predictive distribution $\mathcal{P}(y|x+x';\theta)$, their KL divergence decreases. Since the risk difference $|\Delta Risk|$ can be upper-bounded by the KL divergence, a smaller KL implies a tighter upper bound on the risk change. This indicates that the model exhibits smaller differences in predictive risk between clean and noisy data, meaning its sensitivity to distributional perturbations is reduced, thereby enhancing robustness.

\subsection{Datasets}
We conducted experiments using the following two datasets to evaluate our proposed methods. Table~\ref{tab:dataset} shows the statistics of datasets, and more details of the dataset description are as follows:
\begin{itemize}[leftmargin=15pt, itemsep=2pt, topsep=0pt, parsep=0pt]
    \item \textbf{Assistment2009(ASSIST09)}\footnote{https://sites.google.com/site/assistmentsdata/home/2009-2010-assistment-data/skill-builder-data-2009-2010} is a rich collection of educational data derived from the online learning platform ASSISTments. This dataset spans the academic year 2009-2010 and has been widely used in the CD task.
    
    \item \textbf{SLP-Math}\footnote{https://aic-fe.bnu.edu.cn/en/data/index.html} is a publicly available benchmark dataset derived from the Smart Learning Partner (SLP) online platform. SLP collects learning data intentionally from secondary school students across multiple subjects, providing diverse content. SLP-Math specifically focuses on the mathematics subset of these data.
\end{itemize}

\begin{table}[t]
\centering
\caption{The statistics of two widely used datasets.}
\label{tab:dataset}
\begin{tabular}{c|cccc}
\toprule
Datasets & Students & Exercises & Concepts & Interactions                    \\
\midrule
ASSIST09                    & 4.1k        & 17.7k         & 123   & 324.5k   \\
SLP-Math                   & 1.5k        & 0.9k         & 34     & 57.2k   \\
\bottomrule
\end{tabular}
\end{table}

\subsection{Baseline Methods}
To validate the effectiveness of our proposed OSCD, we conducted extensive experiments and compared several state-of-the-art baselines. The detailed description of these methods is as follows:
\begin{itemize}[leftmargin=15pt, itemsep=2pt, topsep=0pt, parsep=0pt]
    \item \textbf{DINA} employs a deterministic inputs, noisy "AND" gate mechanism to establish an association model between learners' latent abilities and observed outcomes through the mapping relationship between latent attribute vectors and item responses.

    \item \textbf{IRT} is a widely used cognitive diagnosis method. It employs a linear function to model student ability profiles and exercise characteristics unidimensionally.

    \item \textbf{MIRT} extends the IRT framework multidimensionally, modeling both learner traits and exercise features across multiple latent dimensions.
    
    \item \textbf{MF} is a latent factor factorization model. It decomposes the student-exercise score matrix to obtain latent feature representations, thereby enabling exercise performance prediction.

    \item \textbf{NCD} innovatively constructs a neural cognitive diagnosis system that maps learners and exercises into a latent space, employing deep neural networks with monotonicity constraints to model their nonlinear interactions, thereby achieving precise assessment of students' knowledge mastery.
    
    \item \textbf{RCD} innovatively employs graph-structured modeling by constructing a triple relational network comprising student-exercise interaction graphs, concept-exercise association graphs, and concept dependency graphs, while designing multi-level attention mechanisms to achieve co-optimization of node features and graph structures, thereby enhancing the predictive performance of cognitive diagnosis.

    \item \textbf{KSCD} designs a knowledge-sensed cognitive diagnosis approach that maps students, exercises, and interactive/non-interactive concepts into a unified vector space, and generates dynamic mastery vectors through matrix operations, thereby uncovering latent knowledge state association patterns from learning behavior data.

    \item \textbf{KaNCD} is the extension of NCD. It introduces an explicit knowledge association modeling mechanism that infers mastery levels of uncovered knowledge concepts by mining latent dependency relationships among concepts, significantly enhancing the interpretability of diagnostic results.

    \item \textbf{ReliCD} incorporates a bayesian uncertainty quantification mechanism that explicitly models knowledge state uncertainty through personalized prior distributions, while optimizing confidence estimation via calibration loss functions, effectively addressing overconfidence issues in diagnosis caused by data noise and sparsity.

    \item \textbf{ORCDF} explicitly encodes raw response signals as graph-structured features and designs a response-aware graph convolution network to mitigate feature over-smoothing issues in traditional cognitive diagnosis models, significantly improving both prediction accuracy and model interpretability.

    \item \textbf{EMO-NAS-CD} innovatively reformulates neural architecture search as a multi-objective optimization problem, automatically discovering cognitive diagnosis architectures that achieve both high performance and high interpretability through joint optimization of novel interpretability metrics and model efficiency objectives.
\end{itemize}

\end{document}